\documentclass[12pt]{iopart} % for single column layout
\usepackage[utf8]{inputenc}
\usepackage[english]{babel}
\usepackage{footmisc}
\expandafter\let\csname equation*\endcsname\relax
\expandafter\let\csname endequation*\endcsname\relax
\usepackage{amsmath}
\usepackage{cases}
\usepackage{graphicx}
\usepackage{upgreek} 
\usepackage{stmaryrd}
\usepackage{caption}
\usepackage{subcaption}
\usepackage{hhline}
\usepackage{url}
\usepackage{cite}
\usepackage{hyperref}
\usepackage{dblfloatfix} 
\usepackage{enumitem} 
\usepackage{cuted}
\usepackage{ulem} % Needed for \sout

\usepackage{todonotes}

\graphicspath{ {./}}%{./Figures/} } % TODO: We cannot have a folder with figures for submission. All figures in same folder as latex files

\newcommand{\N}[2]{\text{N}\ensuremath{_{#1}^{#2}}}

\renewcommand{\O}[2]{\text{O}\ensuremath{_{#1}^{#2}}}

\newcommand{\p}[1]{\partial_{#1}}

\begin{document}

\title[Estimating streamer properties from measurable parameters]{Estimating the properties of single positive air streamers from measurable parameters}
\author{Dennis Bouwman$^1$, Hani Francisco$^1$,  Ute Ebert$^{1,2}$}
\address{$^1$ Centrum Wiskunde \& Informatica (CWI), Amsterdam, The Netherlands}
\address{$^2$ Department of Applied Physics, Eindhoven University of Technology, PO Box 513, 5600 MB Eindhoven, The Netherlands}

\eads{\mailto{Dennis.Bouwman@cwi.nl}}

\begin{abstract}
We develop an axial model for single steadily propagating positive streamers in air. It uses observable parameters to estimate quantities that are difficult to measure. More specifically, for given velocity, radius, length and applied background field, our model approximates the ionization density, the maximal electric field, the channel electric field, and the width of the charge layer. These parameters determine the primary excitations of molecules and the internal currents. Our approach is to first analytically approximate electron dynamics and electric fields in different regions of a uniformly-translating streamer head, then we match the solutions on the boundaries of the different regions to model the streamer as a whole, and we use conservation laws to determine unknown quantities. We find good agreement with numerical simulations for a range of streamer lengths and background electric fields, even if they do not propagate in a steady manner. Therefore quantities that are difficult to access experimentally can be estimated from more easily measurable quantities and our approximations. The theoretical approximations also form a stepping stone towards efficient axial multi-streamer models.
\end{abstract}


%\submitto
{Submitted to \PSST state of \today}
\maketitle
\ioptwocol % for two column layout

\section{Introduction}

\subsection{The challenge of model reduction}

Streamer discharges occur widely in nature and technology \cite{Nijdam2020ThePhenomena}. The most commonly encountered and studied streamers appear in air and carry positive net charge at their heads. They are the topic of the present study.

The inner structure of a streamer consists of a thin moving curved space charge layer around a weakly ionized channel with strong field enhancement and steep electron density gradients at the tip. This is challenging to simulate numerically, even for a single axisymmetric streamer in a long gap and a low background electric field \cite{Bagheri2018a}. On the other hand many discharge phenomena consist of numerous interacting streamers \cite{Nijdam2020ThePhenomena, Cummer2006SubmillisecondStructure, Kochkin2012ExperimentalAir, Hare2020RadioSteps, Sterpka2021TheRevealed}. This poses a strong motivation to reduce the model while not giving up the physical basis and the model validation achieved in recent years \cite{Bagheri2018a, Li2021ComparingValidation, Dijcks2023ImagingStructures, wang2022quantitative}. 

Streamer discharges consist of clearly distinguishable regions where different physical mechanisms are dominating the behaviour: (i) a non-ionized outer area where the electrostatic Poisson equation has to be solved, (ii) the avalanche zone where photoionization creates many growing electron avalanches, (iii) the moving streamer heads with an active space charge layer where ionization increases rapidly and the field is highest, and (iv) ionized channels with charges and currents and dynamically changing conductivity. Since the regions are governed by different mechanisms we will analyse them separately. Later we match the different regions at their boundaries.

For the channel region an axial approximation has been formulated  in \cite{Luque2014GrowingStreamers,Luque2017StreamerChannels}, but for the streamer head the problem is open. In this work we will concentrate on the heads. To allow for comparison between numerical simulations of the fluid model and analytical approximations, we constrain the analysis to single streamers in a uniform field and mostly to steady propagation. 

\subsection{Steady streamers as a test case}

In sufficiently low electric fields a streamer can propagate at a constant velocity without changing shape \cite{Francisco2021, Francisco2021SimulationsField, Li2022AMixtures, Guo2022AAir}. %This occurs when an unstable balance is reached between growth at the head and loss of conductivity in the channel. 
Such streamers leave no charge behind and their channel electric fields decay back to the applied background field. From now on we will refer to these as \textit{steady}, because in a co-moving coordinate system such uniformly translating streamers are in a steady state. The properties of positive steady streamers can be considered extreme, with velocities as low as \mbox{$3\cdot 10^4$~m/s}, electric fields enhanced to values as high as 222~kV/cm, steep gradients and a strongly curved thin charge layer \cite{Li2022AMixtures}.\\

The analysis presented in this work focuses heavily on steady streamers, since it is mathematically convenient to consider steady state solutions, as they have no explicit time dependence in a co-moving frame. Furthermore, we validate our approximations by comparing them to simulated results of a steady streamer. It must be noted that such a steady state approach could also be considered for accelerating streamers, since their properties typically change slowly with respect to other relevant time scales \cite{Pavan2020}. To that end we also compare our approximations to simulations of three accelerating streamers.

\subsection{Earlier work}
A classical challenge is to develop equations of motion where the head is characterized by a few numbers like radius $R$ and velocity $v$.  One of the first proposed analytic relations between $R$ and $v$ date back to 1965 \cite{Loeb1965IonizingGradient} and an `order-of-magnitude' model for the parameters of streamers was given in 1988 \cite{Dyakonov1988TheorySemiconductors, Dyakonov1989}. A later experimental investigation proposed a data fit where the velocity depends on the radius squared, i.e.\ \mbox{$v\sim R^2$} \cite{Briels2008PositiveEnergy}, and in \cite{Naidis2009} an approximate relation based on \cite{Loeb1965IonizingGradient} was proposed where the velocity is also a function of the maximum electric field at the tip $E_{\rm max}$, i.e.\ \mbox{$v=v(R, E_{\rm max})$}. Other important theoretical results are: an approximation for the ionization density \cite{Dyakonov1989, Naidis1997, Li2007DeviationsHeads}, energy efficiency estimates for radical production \cite{Naidis2012}, an analytic investigation of the avalanche zone dynamics \cite{Pancheshnyi2001}, 1.5D models that require a prescribed radius \cite{Pavan2020, Lehtinen2021} and an estimate for charge layer width based on the notion of an effective ionization length \cite{Niknezhad2021UnderlyingStreamers}. An application of streamer theory is to infer difficult-to-measure properties, such as $E_{\rm max}$, from measurable parameters. For example, in \cite{Babaeva2021UniversalMedia} the authors estimate a parameter range for $E_{\rm max}$ on the basis of observed radius and velocity. Another example is \cite{Pancheshnyi2001}, where an analysis of the avalanche zone gives an approximate relation between $R$, $E_{\rm max}$ and the head potential.

These theoretical results have improved our understanding of streamer dynamics and illustrated complex relations between different parameters. However, some ideas proposed in earlier works fail to agree with results from numerical simulations. At several instances throughout this work we will provide an in-depth evaluation of earlier work and propose improvements.\\

\subsection{Content of the paper}
In this work we develop semi-analytic approximations for the fluid model of single positive streamers that estimate difficult-to-measure quantities based on observable parameters. Specifically, we will show how velocity $v$, radius of curvature $R$, length $L$ and background field $E_{\rm bg}$ determine ionization density $n_{i, \rm ch}$, charge layer width $\ell$ and the maximum and channel electric fields $E_{\rm max}$ and $E_{\rm ch}$, respectively:
\begin{equation}\label{eq:problem_formulation}
    (v,\, R,\, L,\, E_{\rm bg}) \to (n_{i, \rm ch},\, \ell,\, E_{\rm max},\, E_{\rm ch})
\end{equation}
The derivation of our model starts by first defining different regions where specific physical mechanisms dominate. Then we provide analytic approximations for each of these regions separately. Finally we match the different regions at their boundaries and implement a self-consistent solution method. This results in a self-contained axial model which agrees well with numerical simulations. This means that our framework can complement experimental measurements when important streamer characteristics, i.e.\ the parameters on the right-hand side of equation \eqref{eq:problem_formulation}, are difficult to measure precisely.\\

In section \ref{sec:model} we outline the classical fluid streamer model and the numerical implementation used for axisymmetric simulations. Furthermore, we discuss the results of numerical simulations in detail and introduce important definitions and conventions.  In section \ref{sec:electrondynamics} we integrate through the charge layer and obtain an analytic formula for the ionization density. In section \ref{sec:avalanchezone} we give an analysis of the electron avalanche dynamics in the region ahead of the streamer. In section \ref{sec:head_potential} we explore the notion of the streamer head potential. In section \ref{sec:solutionandresults} we describe our solution method and validate our approximations with numerical results of the fluid model.

\section{Model description, definitions and conventions}\label{sec:model}

In this section we will present the classical fluid model for positive streamers in air at standard temperature and pressure. We discuss the numerical implementation, used to obtain reference solutions in homogeneous background electric fields below the breakdown value. The same implementation was used in \cite{Francisco2021, Francisco2021SimulationsField} to study steady streamers. Furthermore, we will also give definitions of macroscopic parameters and clarify other conventions and terminology. 

\subsection{Description of the model}

\subsubsection{Basic equations. $\quad$}

We employ the classic fluid streamer model with local field approximation and without ion mobility. 
We only account for two charged species: the electron density $n_e$ and the net ion density \mbox{$n_i=n_{+} - n_{-}$}, with $n_{\pm}$ denoting the number density
of all positive or negative ions. One can use just one ion density $n_i$ instead of several ion species in regions where ion drift, electron detachment, and electron ion recombination can be neglected, as is the case in the streamer head. The electron density evolves according to a drift-diffusion-reaction equation while ions are considered immobile
\begin{align}
        \p{t} n_e =&  \nabla \cdot (\mu n_e {\bf E} + D\nabla n_e) \label{eq:electron_transport}
          + S_i + S_{ph},\\
        \p{t} n_i  =&  S_i + S_{ph}, \label{eq:pos_ion_transport}
\end{align}
with ${\bf E}$ the electric field, $\mu(E)$ the electron mobility, and $D(E)$ the electron diffusion coefficient. $S_i$ and $S_{ph}$ are the source terms for the effective impact ionization and photo-ionization respectively. We neglect electron diffusion, which is typically a good approximation except in low $E_{\rm bg}$ where we have steep gradients in the charge layer. Effective impact ionization is given by
\begin{equation} \label{eq:Si}
    S_i = |{\bf j}_e|\alpha_{\rm eff}, 
\end{equation}
where \mbox{${\bf j}_e=-\mu n_e {\bf E}$} is the drift current density of electrons, \mbox{${\bf j} = -e{\bf j}_e$} is the electric current density, $e$ is the elementary charge, and $\alpha_{\rm eff}(E)$ is the effective ionization coefficient. The data for the transport and reaction coefficients are discussed in the next section. The photo-ionization source term in a volume $V$ is given by
\begin{equation}\label{eq:sim_photoionizationsourceterm}
    S_{ph}({\bf r}) = \iiint_V \frac{I({\bf r'})f(|{\bf r- \bf r'}|)}{4\pi|{\bf r-\bf r'}|^2}d^3r',
\end{equation}
with $I({\bf r'})$ the source of ionizing photons, $f(r)$ the absorption function and $4\pi|{\bf r-\bf r'}|^2$ is a geometric factor. Following Zheleznyak's model \cite{Zhelezniak1982PhotoionizationDischarge, Pancheshnyi2014Photoionization2}, $I({\bf r})$ can be expressed as
\begin{equation}\label{eq:sim_photonsource}
    I({\bf r}) = \frac{p_q}{p+p_q}\xi S_i({\bf r}),
\end{equation}
with $p$ the actual pressure, \mbox{$p_q=40$~mbar} the quenching pressure of the gas-mixture, and \mbox{$\xi=0.075$} a proportionality factor relating impact excitation to impact ionization $S_i$. 
The absorption function $f(r)$ is given by
\begin{equation}
    f(r) = \frac{\exp(-\chi_{\rm min}p_{O_2}r)-\exp(-\chi_{\rm max}p_{O_2}r)}{r\ln(\chi_{\rm max}/\chi_{\rm min})},
\end{equation}
with $\chi_{\rm max}=150/$~(mm bar), $\chi_{\rm min}=2.6/$~(mm bar), and $p_{O_2}$ is the partial pressure of oxygen. For air at 300~K and 1~bar, the corresponding absorption lengths are 33~$\mu$m and 1.9~mm. 

The electric field follows from Poisson's equation for the electric potential $\phi$
\begin{align}
    \epsilon_0\nabla^2\phi &= -e n_q,\label{eq:poisson_equation}\\
    {\bf E} &= -\nabla\phi,\label{eq:electric_potential}
\end{align}
with $\epsilon_0$ the dielectric constant, $e$ the elementary charge, and \mbox{$n_q = n_i - n_e$} the charge number density.

\subsubsection{Implementation of axisymmetric simulations. $\quad$}
In this work we compare our axial analytical approximations with the axisymmetric solutions of equations \eqref{eq:pos_ion_transport}-\eqref{eq:electric_potential} obtained by numerical simulation. The numerical model uses the afivo-streamer code \cite{Teunissen2017SimulatingFramework, Teunissen2018Afivo:Methods}. The computational setup is the same as in earlier studies \cite{Francisco2021, Francisco2021SimulationsField} to which we refer for an in-depth discussion.\\

The transport and reaction coefficients are calculated by Bolsig+ \cite{Hagelaar2005} (version 12/2019) using cross sections from the Phelps database~\cite{phelps_data} under the assumption that the evolution of the electron density follows an exponential temporal growth or decay \cite{Wang2022AAirb}. We use the same data for the analytical and the  numerical models. Additionally, the numerical model for the axisymmetric simulations uses continuity equations for a number of species such as \O{2}{+}, \O{2}{-}, \N{2}{+}, \N{4}{+}, \textit{etc.} as listed in \cite{Francisco2021SimulationsField}. This more extended plasma chemistry model helps stabilizing the steady streamer at the lowest background electric field, and it is consistent with the two-species model for $n_e$ and $n_i$ in the streamer head, as recalled above.

The photo-ionization integral in equation \eqref{eq:sim_photoionizationsourceterm} is approximated by a set of Helmholtz differential equations with Bourdon's three-term approximation \cite{Bourdon2007EfficientEquations}. This approximation introduces small changes the photon absorption lengths. However, in \cite{Wang2022AAirb, Bagheri2019} it was shown this has essentially no measurable influence on streamer discharge propagation in air.

\subsubsection*{Computational domain. $\quad$ }
The computational domain consists of a cylinder with 40~mm length and 20~mm radius, and planar electrodes on top and below. We impose cylindrical symmetry for domain and streamers; and we call the longitudinal coordinate $\zeta$, and the radial coordinate $r$. An electric field is applied in the $\zeta$-direction by fixing an electric potential difference between the electrodes. We use homogeneous Neumann boundary conditions for the potential in the $r$-direction, which means that the electric field is parallel to the lateral boundary. Homogeneous Neumann boundary conditions are also used for the electron density on all boundaries.

\subsubsection*{Initial conditions. $\quad$}
For the initiation of a streamer discharge, we placed two neutral seeds composed of electrons and positive ions at the upper boundary of the domain. The uppermost seed creates a region of field enhancement, and the seed below it supplies the initial electrons ahead of the forming streamer, before photo-ionization sets in. More details on the seeds --- their densities, coordinates, and sizes --- can be found in \cite{Francisco2021, Francisco2021SimulationsField}.\\

In low electric fields, an initial transient electric field is needed to ensure the inception of a streamer discharge. In this research we will consider homogeneous background electric fields from 4.5 to 24~kV/cm, all below the breakdown value of 28~kV/cm. At 4.5~kV/cm, a streamer propagating at constant velocity and shape was obtained using the velocity control method \cite{Li2022AMixtures}. At 10~kV/cm we adopt the same initial transient electric field as discussed in \cite{Francisco2021SimulationsField}. 

To accommodate for the relatively small size of the steady streamer we used a grid with a minimum cell width of 0.6 $\mu$m. For the accelerating streamers, the mesh refinement routines are identical to those in \cite{Francisco2021SimulationsField}.
% fixed the computational grid instead of employing adaptive mesh refinement in this simulation. From the axis of propagation up to 0.5~mm radially outwards, the grid cells were fixed to have a width of 1~$\mu$m.

\begin{figure*}
    \centering
    \includegraphics{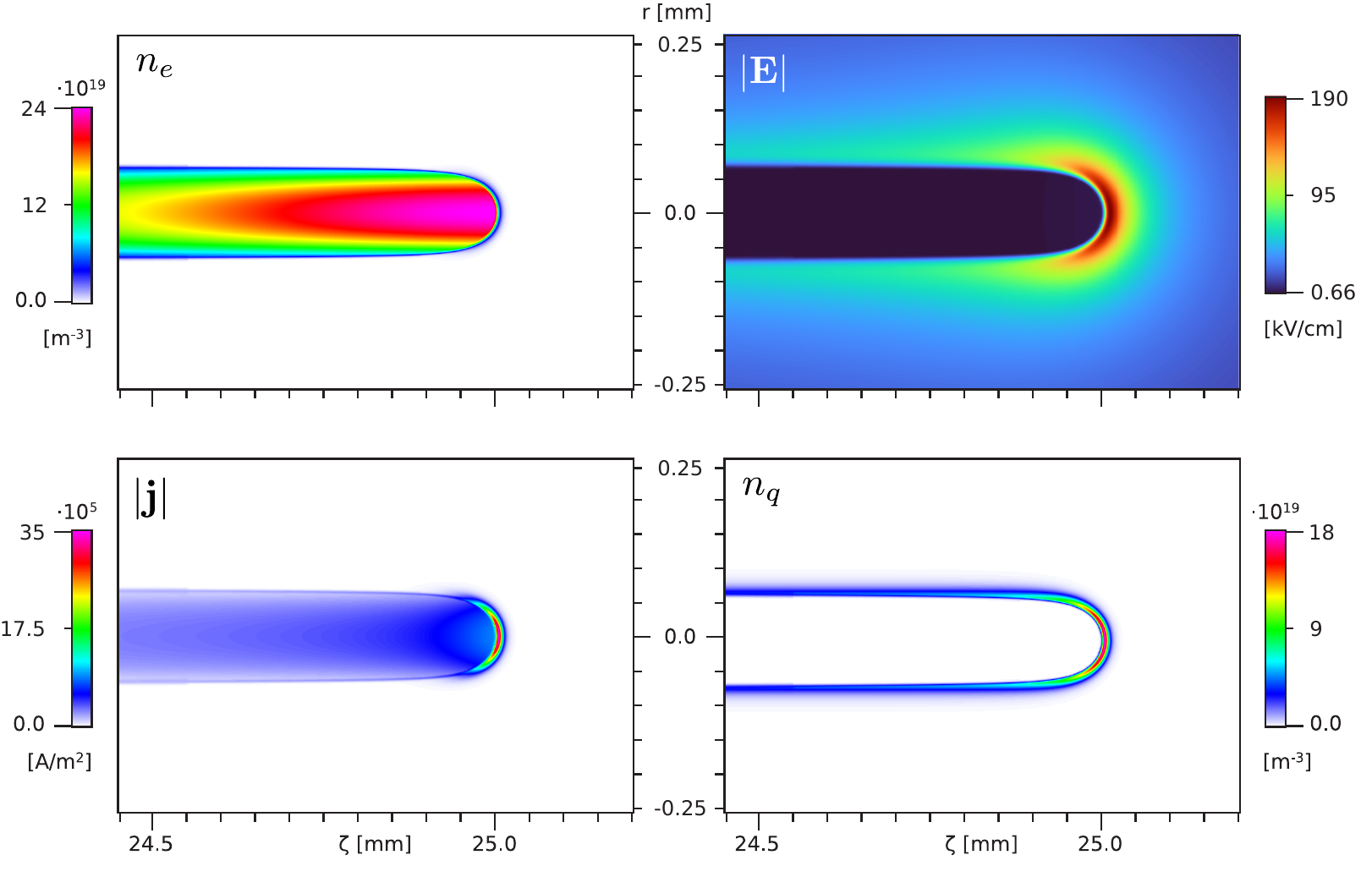}
    \caption{Electron density $n_e$, electric field strength $|\bf E|$, strength of the electric current density $|\bf j|$ and charge number density $n_q$ of a steadily propagating streamer in a background field $E_{\rm bg}$ of $4.5~$kV/cm. The figure zooms into the area around the streamer head.}
    \label{fig:numericalsimulation}
\end{figure*}

\subsection{Description of axisymmetric simulation results}

\subsubsection{The steady streamer in detail. $\quad$}

In this section we will discuss one of these simulations in detail, the steady streamer at a background electric field of 4.5~kV/cm. We recall that a steady streamer \cite{Francisco2021SimulationsField, Francisco2021, Li2022AMixtures, Guo2022AAir} looses its conductivity at its back end due to electron attachment and electron ion recombination, that it leaves no electric charge behind, but carries a fixed amount of charge along, and that it propagates with constant velocity and shape. Figure~\ref{fig:numericalsimulation} zooms into the front part of this streamer and shows four important quantities: the electron density $n_e$, the magnitude $|\bf E|$ of the electric field, the magnitude $|\bf j|$ of the electric current density and the charge number density $n_q$. From these quantities we can distinguish three regions with different dynamics:
\begin{enumerate}
    \item \textit{The channel }is the conductive interior of the streamer. We have a high electron density here and the plasma is quasi-neutral, \mbox{$n_q\approx 0$}. The electron density in the low axial electric field gives rise to an electric current flowing along the channel.
    \item \textit{The charge layer }is a layer of (positive) charge which surrounds and partially screens the channel. At the streamer head, the curvature of the charge layer leads to high electric field enhancement ahead of the front. In fact, we find the maximum electric field $E_{\rm max}$ here, with its location denoted by $\zeta_{\rm tip}$. As the electron density is high as well, we here have a high impact ionization rate and large currents resulting in the movement of the streamer head. The charge layer in the streamer head is also referred to as the ionization front. The width $\ell$ of the charge layer is much smaller than its radius $R$ of curvature; this is required for the strong field enhancement ahead of the layer.
    \item \textit{The avalanche zone }of a positive streamer is the region ahead of the charge layer, so the electric charges in this region have a negligible effect on the electric field distribution. Without photo-ionization or some background electron density it could be neglected, but for positive streamers in air the photo-ionization creates many growing electron avalanches moving towards the charge layer. Close to the layer there is a high electric field, which means that a significant electron current is created which maintains the active ionization front. Specifically in air without background ionization, the electron density vanishes with an asymptotic decay dictated by photon absorption \cite{Bouwman20223DCombustion}
\begin{align}
    n_e(\zeta) &\propto \zeta^{-1}e^{-k\zeta} \mbox{ with }k=\chi_{\rm min}p_{\O{2}{}},\nonumber\\
    &\text{ for } \zeta\gg\zeta_{\rm tip}.
\end{align}
\end{enumerate}

\begin{figure}
    \centering
    \includegraphics[width=0.5\textwidth]{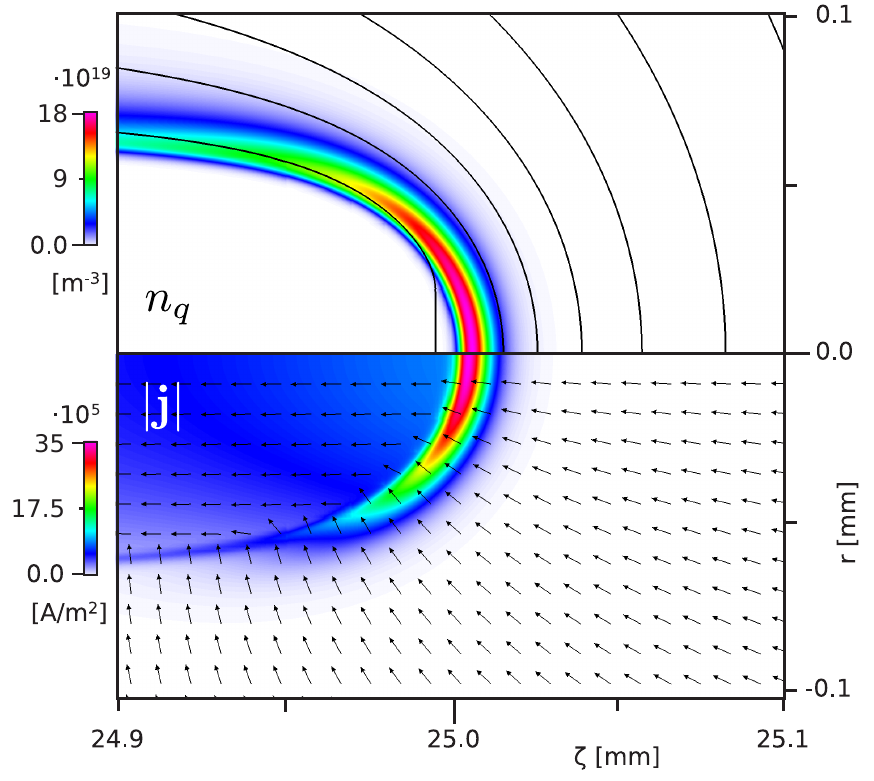}
    \caption{Charge number density $n_q$ and magnitude of the electric current density $|\bf j|$ of the steady streamer in a field of $4.5~$kV/cm in color-coding. In the upper half of the plot, equipotential lines are laid over the charge number density. In the lower half, the arrows show the normalized direction of the electron drift ${\bf j}_e = -{\bf j}/e$.}
    \label{fig:electric_field_lines}
\end{figure}
\subsubsection{Directions of currents and fields. $\quad$}

In figure \ref{fig:electric_field_lines} we zoom further into the ionization front and highlight important geometric features. We show the charge number density $n_q$ and the magnitude of the electric current density $|\bf j|$ again in color-coding, but additionally we have visualized the direction of the current density by normalized arrows in the lower half of the plot, and the equipotential lines in the upper half of the plot. Note that in the avalanche zone the direction of the electron current ${\bf j}_e$ is radially inwards in a nearly spherical geometry, whereas in the channel the electron drift is homogeneously directed backwards parallel to the axis of propagation. Furthermore, the equipotential lines are very well aligned with the charge layer. This means that the electric current is essentially perpendicular to the layer in this region.

\begin{figure*}[htp]
\centering
\begin{subfigure}{.33\textwidth}
  \centering
  \includegraphics[width=\linewidth]{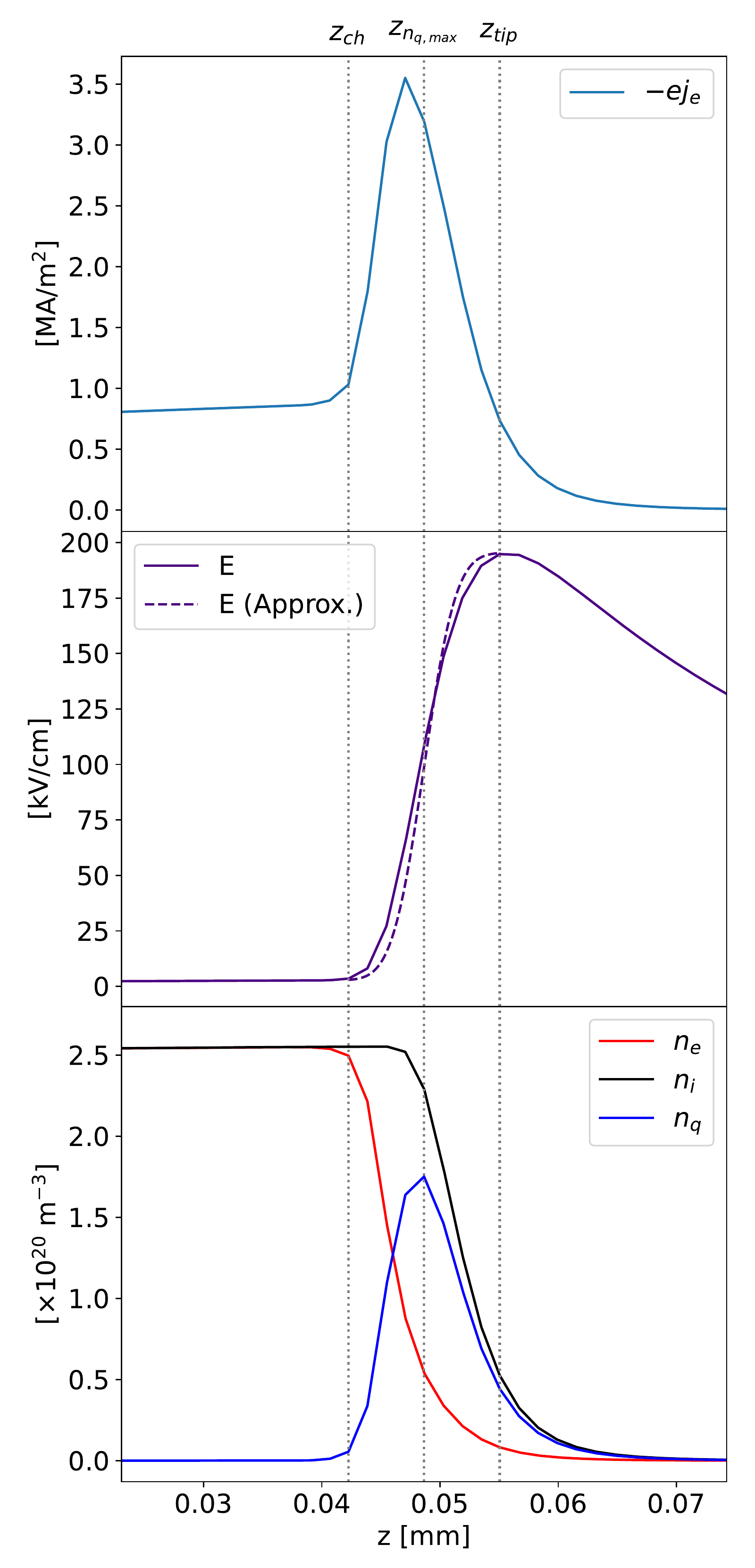}
  \caption{4.5 kV/cm}
  \label{fig:sub1}
\end{subfigure}%
\begin{subfigure}{.33\textwidth}
  \centering
  \includegraphics[width=\linewidth]{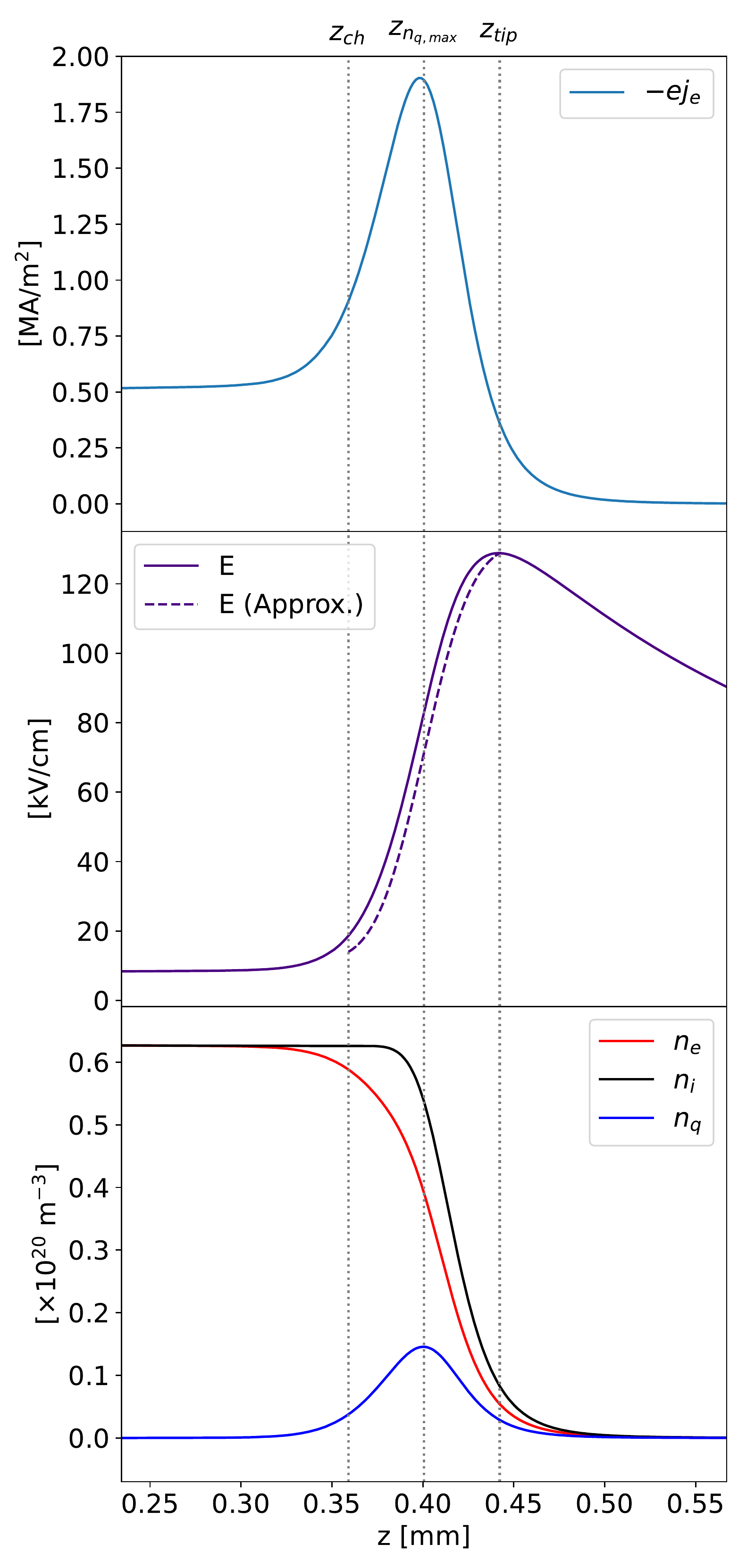}
  \caption{14 kV/cm}
  \label{fig:sub1}
\end{subfigure}%
\begin{subfigure}{.33\textwidth}
  \centering
  \includegraphics[width=\linewidth]{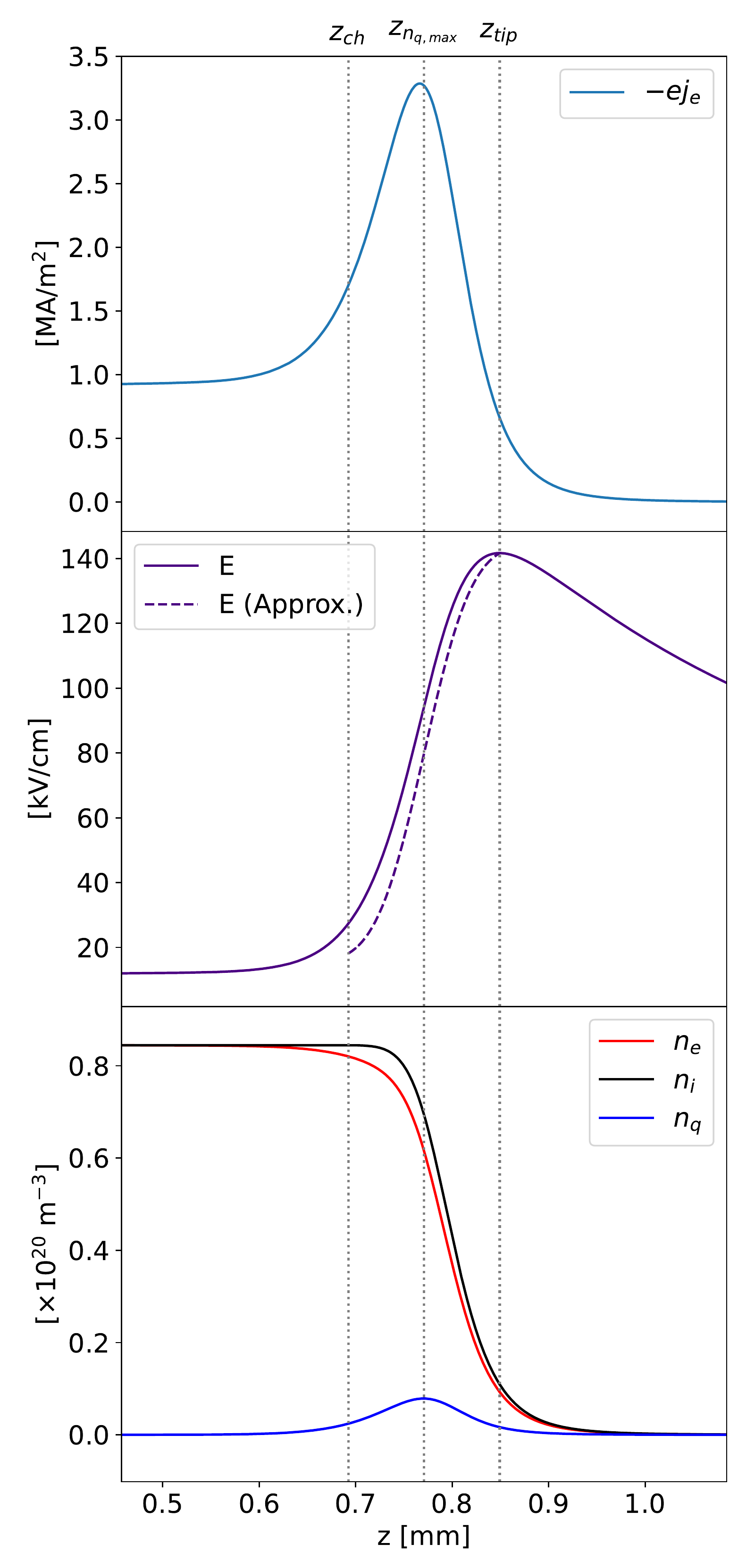}
  \caption{24 kV/cm}
  \label{fig:sub1}
\end{subfigure}%
\caption{Current density, electric field and particle densities on axis for streamers in three background electric fields. All streamers are shown when the head has reached \mbox{$\zeta_{\rm tip}=15$}~mm. The origin of the coordinate system, $z=0$, is at the centre of the hemisphere fitted through the maximum of the charge number density. The corresponding $v$ and $R$ are shown in figure \ref{fig:vandR}.}
\label{fig:on_axis_currents}
\end{figure*}

\subsubsection{Streamers in different fields. $\quad$}
In figure \ref{fig:on_axis_currents} we 
show current densities, electric fields and electron and charge densities on the streamer axis, now not only for the steady streamer in the field of 4.5~kV/cm, but also for accelerating streamers in background fields of 14 and 24~kV/cm when the streamer heads reached \mbox{$\zeta=15$~mm}. In more detail, the upper plots show the electric current density. The middle plots show the electric field (solid line) with our approximation (dashed line) of section~\ref{sec:ionization_front}, and the lower plots show $n_e$, $n_i$ and $n_q$. %We also annotate the location of the maximum of the charge number density $\zeta_{n_{q, \rm max}}$, of the maximum of the electric field $\zeta_{\rm tip}$, and of the start of the channel $\zeta_{\rm ch}$.

\subsection{Definitions and conventions} \label{sec:definitions_and_conventions}
\begin{figure}
    \centering
    \includegraphics[width=0.5\textwidth]{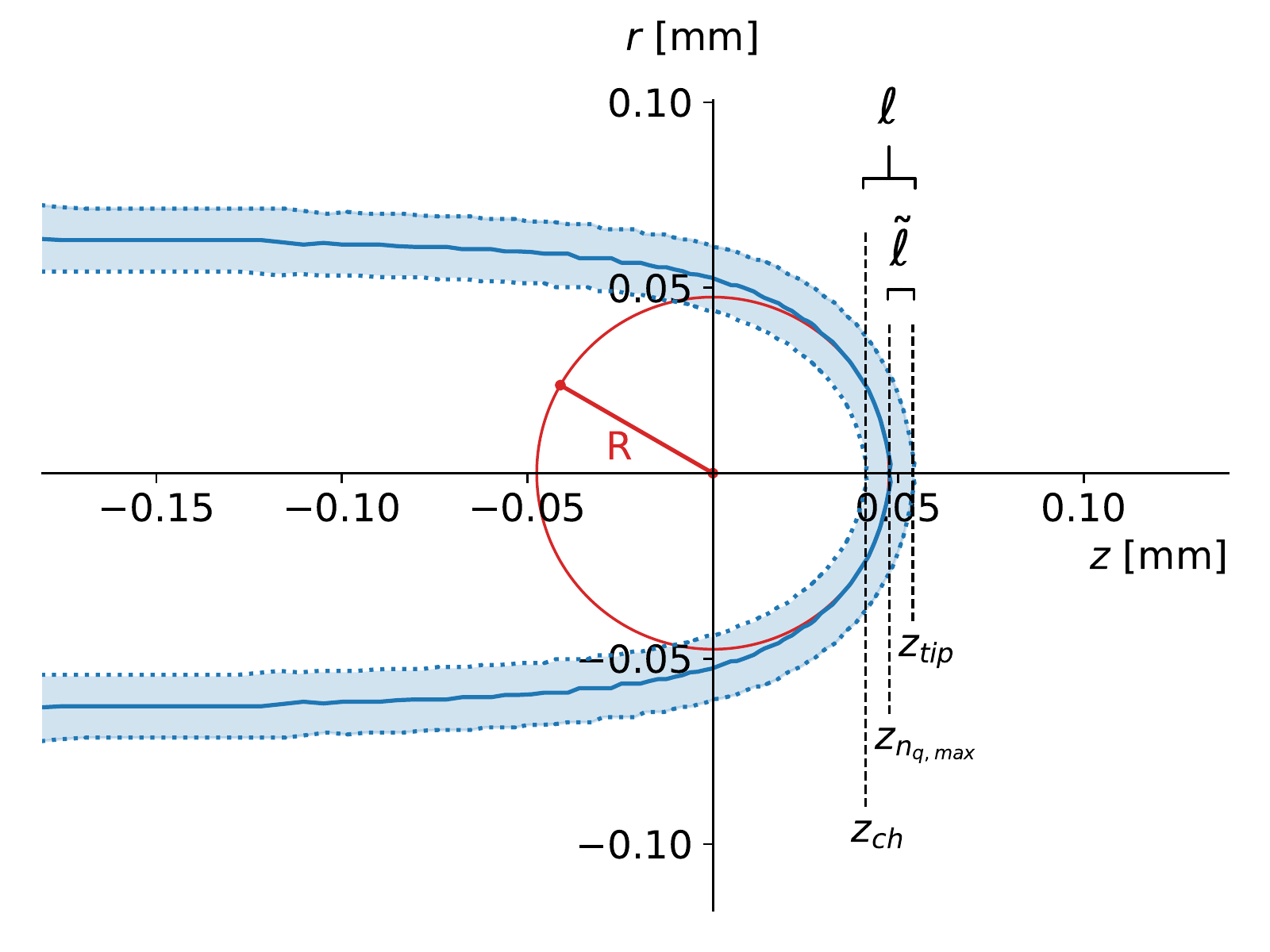}
    \caption{The charge layer within the co-moving coordinate system $(r,z)$ at \mbox{$E_{\rm bg} = 4.5$~kV/cm}. The solid blue line represents the maximum of $n_q$ (for each $z$) from numerical simulation and the shaded area is the corresponding charge layer parameterized using $\ell$. Also shown are: the tangent circle with radius $R$, $\tilde \ell$ and the  positions $z_{\rm ch}$, $z_{\rm tip}$ and $z_{n_{q, \rm max}}$. }
    \label{fig:shape_and_grid}
\end{figure}

In this paper we develop an axial model for the dynamics in charge layer and avalanche zone, based on analytical approximations. Here we introduce definitions and conventions for this purpose. A schematic is given in figure \ref{fig:shape_and_grid}.

\begin{figure*}
    \begin{subfigure}{.5\textwidth}
      \centering
      \includegraphics[width=\linewidth]{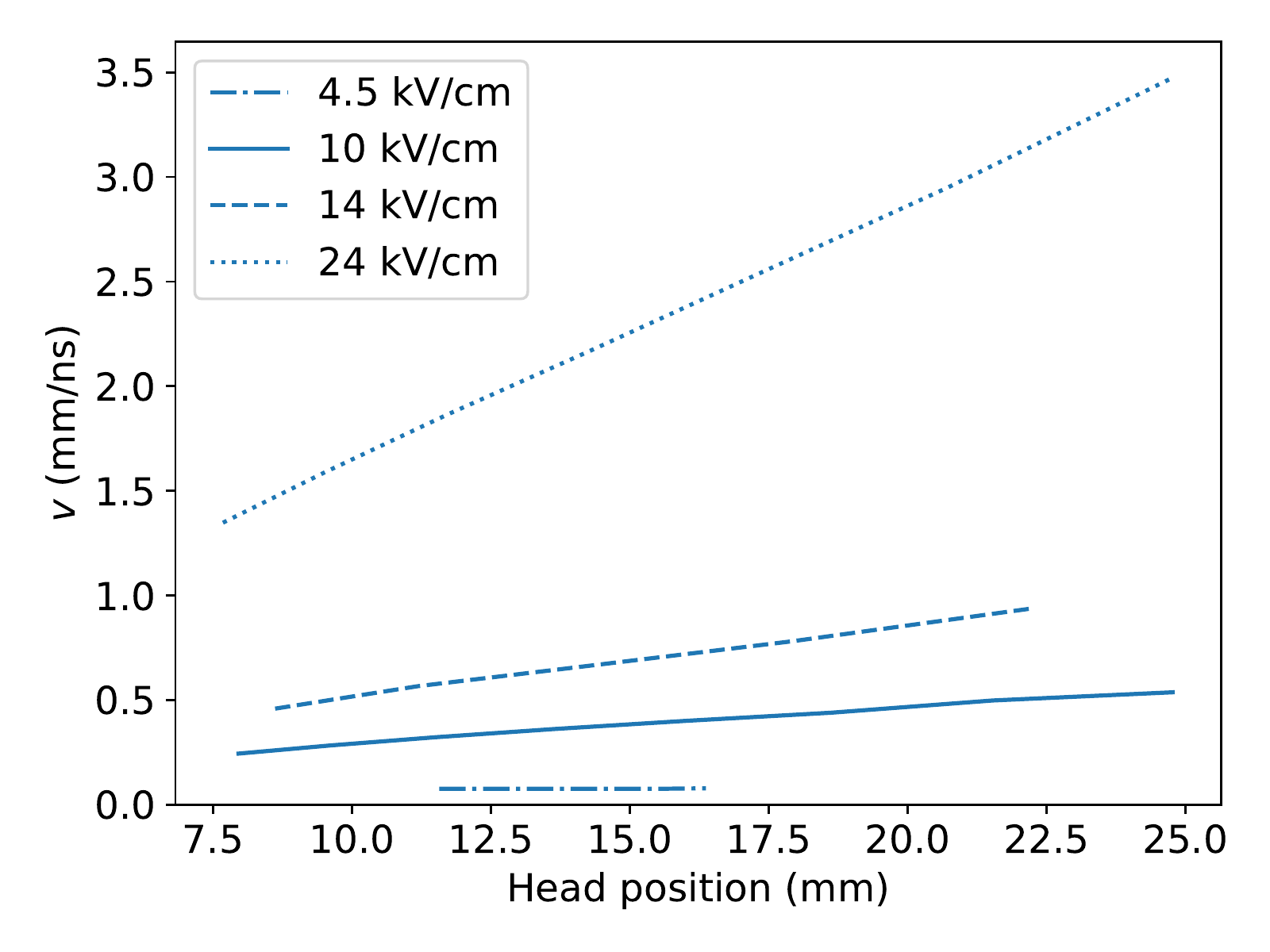}
      \subcaption{Velocity}
      \label{fig:v}
    \end{subfigure}%
        \begin{subfigure}{.5\textwidth}
      \centering
      \includegraphics[width=\linewidth]{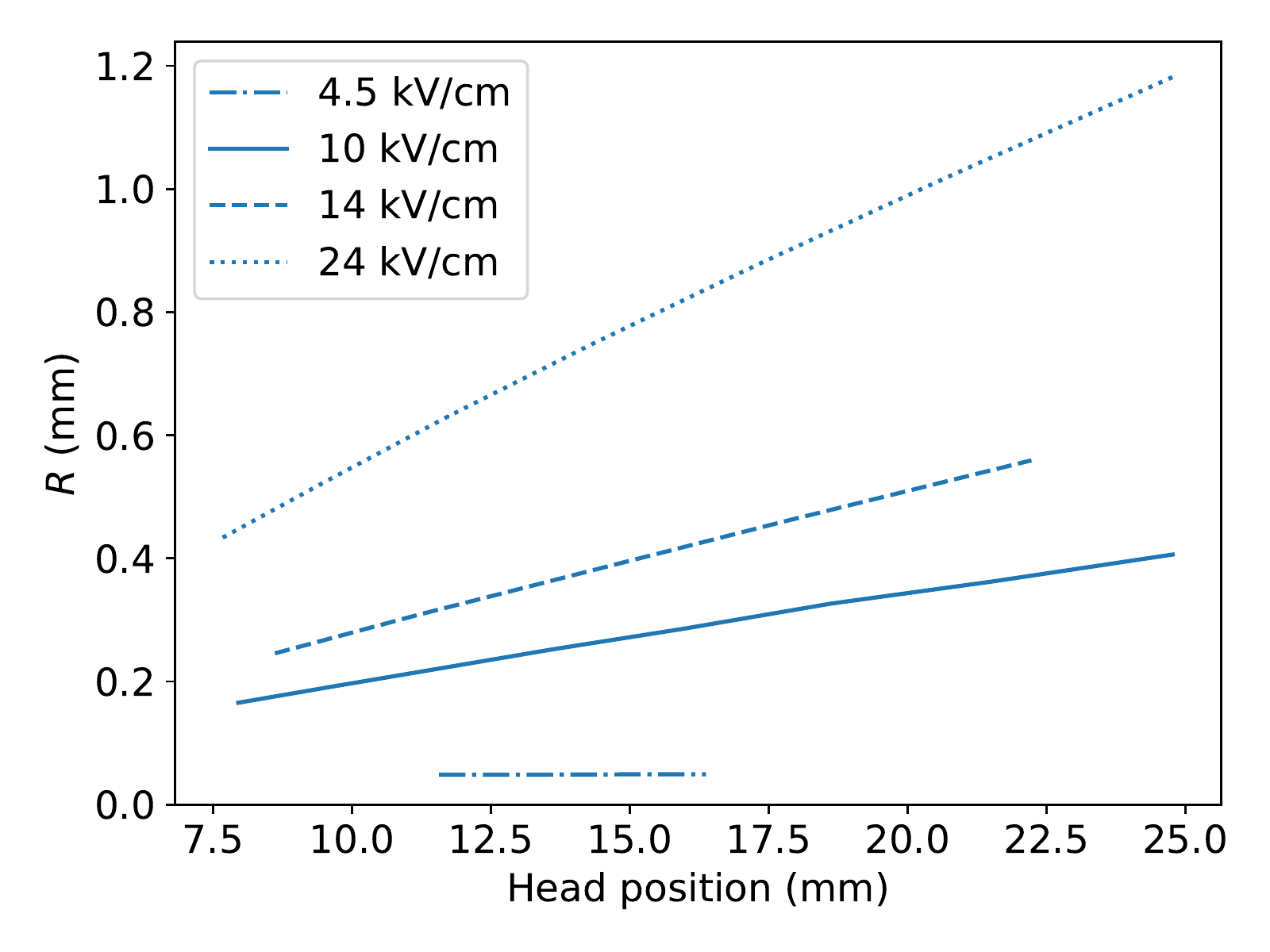}
      \subcaption{Radius of curvature}
      \label{fig:R}
    \end{subfigure}%
    \caption{The velocity and radius as a function of the head position extracted from simulations at different background electric fields.}
    \label{fig:vandR}
\end{figure*}

\subsubsection{Definition of velocity and co-moving coordinate system. $\quad$ }
We define the streamer velocity $v$ as the velocity of the location of the maximal electric field at the streamer tip 
\begin{equation}
    v(t) = \frac{{\rm d}\zeta_{\rm tip}(t)}{{\rm d}t}.
\end{equation}
The velocity extracted from simulations is shown in figure \ref{fig:v}. We introduce a coordinate system $(r,z)$ that moves in the $\zeta$ direction with velocity $v$. The $z$ coordinate can be written as
\begin{equation}
    z=\zeta - vt.
\end{equation}
Temporal derivatives transform to the new coordinate system as
\begin{equation} \label{eq:co-moving1}
    \partial_t\big|_{\zeta} = \partial_t\big|_{z}-v\partial_z,
\end{equation}
where $\partial_t\big|_{z}$ denotes the partial derivative $\p{t}$ in the co-moving frame $(r,z)$. For steady motion we thus can replace
\begin{equation} \label{eq:co-moving2}
    \partial_t\big|_\zeta = -v\partial_z.
\end{equation}
in the co-moving frame $(r,z)$.

\subsubsection{Parameterizing the charge layer. $\quad$ }
We will characterize the charge layer by two maxima, namely the maximum of the electric field and the maximum of the charge number density. On the streamer axis we will denote them as $z_{\rm tip}$ and $z_{n_{q, \rm max}}$, and their distance as
\begin{equation}
    \tilde \ell = z_{\rm tip} - z_{n_{q, \rm max}}
\end{equation}
in the co-moving coordinate system $(r,z)$ defined below. The two maxima are also illustrated in figure \ref{fig:on_axis_currents}. There it is also shown that $z_{n_{q, \rm max}}$ is located roughly in the middle of the charge layer and that $j$ is approximately symmetric in the vicinity of this maximum. Therefore we define the interior boundary $z_{\rm ch}$ of the charge layer as
\begin{equation}
     z_{\rm ch} = z_{\rm tip} - \ell \mbox{ with } \ell = 2\tilde\ell.
\end{equation}

\subsubsection{Definition of radius and of origin of coordinate system. $\quad$}
We will characterize the streamer head by its radius of curvature $R$, defined as the radius of the circle which best approximates the curved charge layer at the streamer tip. This parameter is extracted from simulated data by fitting a semicircle through the maximum, for each $z$, of the charge layer, cf.\ figure \ref{fig:shape_and_grid}. The extracted $R$ is insensitive to fitting parameters provided the region is chosen sufficiently small. We therefore take this region to be \mbox{$[z_{n_{q, \rm max}}-4\tilde \ell, z_{n_{q, \rm max}}]$}. The radius of curvature extracted from simulations is shown in figure \ref{fig:R}.

$R$ is an important quantity because it determines the spatial decay of electric field and currents in the avalanche zone near the charge layer, as can be seen in figure \ref{fig:electric_field_lines}. There the equipotential lines trace the shape of the charge layer sufficiently close to the axis of propagation. We choose the centre of the sphere as the origin of the co-moving coordinate system, $(r,z)=(0,0)$, as shown in Fig.~\ref{fig:shape_and_grid}.

\subsubsection{Definition of streamer length for steady streamers}
The steady positive streamers reported in \cite{Francisco2021, Francisco2021SimulationsField, Li2022AMixtures} are all `detached' from their point of inception. By this we mean that due to attachment and recombination processes the channel looses its conductivity to the point that the streamer cannot be considered as connected to an electrode or initial ionized seed. For these detached streamers it is more useful to characterize streamer length with a typical length scale for the loss of conductivity $L_{\rm loss}$
\begin{equation}
    L_{\rm loss} = v\tau,
\end{equation}
with $\tau$ the electron loss time representing the collective timescale of all conductivity loss processes. The studies \cite{Francisco2021SimulationsField, Li2022AMixtures} contain investigations of $L_{\rm loss}$ and $\tau$. In particular, it is analyzed how these quantities depend on the gas-composition and the electric field.

\section{The charge layer $(z_{\rm ch}\leq z<z_{\rm tip})$}\label{sec:electrondynamics}

In this section we formulate axial approximations for the total current density and for the electron and ion densities in the space charge layer, i.e., in the region between the front end $z_{\rm ch}$ of the channel and the maximum of the electric field $z_{\rm tip}$ (cf figure \ref{fig:shape_and_grid}). The width $\ell$ of this region is much smaller than the radius of curvature $R$, therefore this layer can approximately be treated as planar.

We can neglect photo-ionization $S_{ph}$ in the charge layer since it is much smaller than $S_i$. Photo-ionization only matters in the avalanche zone due to its nonlocality. We also neglect the diffusive current assuming that it is dominated by convection. 

\subsection{Current densities in the charge layer} \label{sec:totalcurrentdensity}
\begin{figure}
    \centering
    \includegraphics[width=0.5\textwidth]{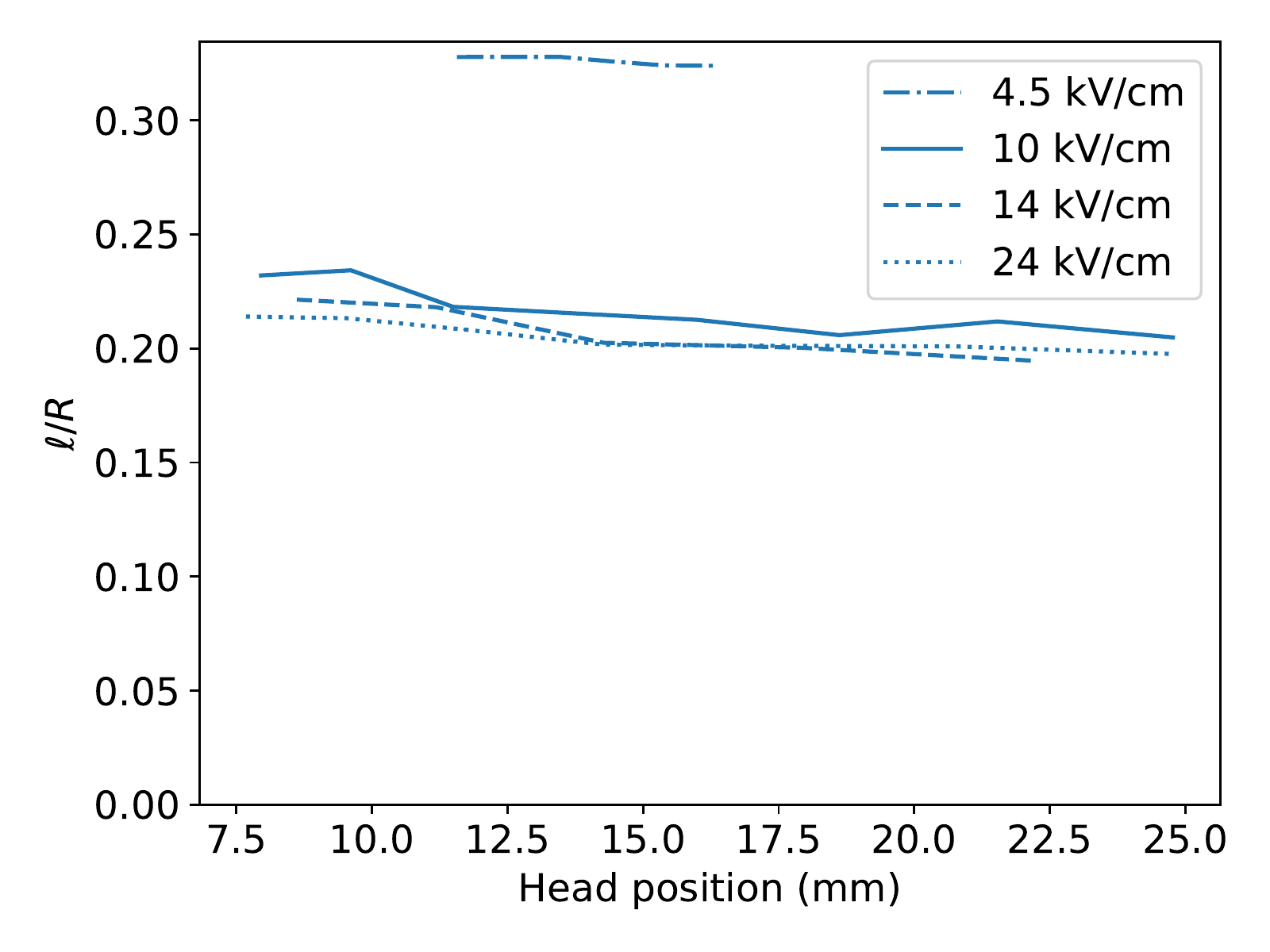}
    \caption{The dimensionless parameter $\ell/R$ as a function of the head position extracted from simulations at different background electric fields. This parameter characterizes the validity of the planar front approximation. The fluctuations observed are due to the small size of $\tilde \ell$ which is only a few times the smallest grid size.}
    \label{fig:l_over_R}
\end{figure}

Due to charge conservation and the Poisson equation of electrostatics, the total current density ${\bf j}_{\rm tot}$ is a conserved quantity
\begin{equation} \label{eq:conservationoftotalcurrent}
    \nabla\cdot {\bf j}_{\rm tot} = 0, \mbox{ where }{\bf j}_{\rm tot} = {\bf j} + \epsilon_0\p{t} {\bf E}.
\end{equation}
For steady motion in a co-moving frame $z$, the total current is ${\bf j}_{\rm tot} = {\bf j} - v\epsilon_0\p{z} {\bf E}$. Note that the current densities are taken in the stationary frame, $\mathbf{j}=-e\mu n_e\mathbf{E}$ expressed as a function of $z$.\\

To solve for ${\bf j}_{\rm tot}$ we approximate the charge layer at the tip as a planar surface. The validity of this approximation is governed by the dimensionless parameter $\ell/R$. More specifically we require \mbox{$\ell/R\ll 1$}, which usually holds for streamers as is shown in figure \ref{fig:l_over_R}. In that case only the $z$-derivative of the divergence operator is non-vanishing. Then, equation \eqref{eq:conservationoftotalcurrent} prescribes that ${\bf j}_{\rm tot}$ is constant. With a boundary condition at $z_{\rm tip}$ this leads to the axial approximation
\begin{equation}\label{eq:J_dependence_on_r}
j_{\rm tot}(z)= j_{\rm tot}(z_{\rm tip}).
\end{equation}
Furthermore, the electric field is maximal at $z_{\rm tip}$, hence \mbox{$\left.\p{z}{\bf E}\right|_{z_{\rm tip}} = 0$} and the displacement current vanishes there
\begin{equation}
    j_{\rm tot}(z_{\rm tip}) = -e j_{e, \rm tip},
\end{equation}
where $j_{e, \rm tip}$ is the electron current density on axis at $z_{\rm tip}$.
Similarly, the displacement current also vanishes approximately in the channel, where the electric field and electron density are nearly constant on-axis. 
This gives us
\begin{equation}\label{eq:eqv}
    j_{e, \rm ch} = j_{e, \rm tip},
\end{equation}
where $j_{e, \rm ch}$ is defined analogously as $j_{e, \rm tip}$. Finally, combining this result with equation \eqref{eq:J_dependence_on_r} determines \mbox{$-j_{\rm tot}/e= j_{e, \rm tip}=j_{e, \rm ch}$}. 

An analysis of the total current density at $z_{\rm ch}$ and $z_{\rm tip}$ was also proposed in \cite{Babaeva1997DynamicsFields}. They held that $j_{e, \rm tip}$ vanishes which would mean that $j_{\rm tot}(z_{\rm tip})$ is completely determined by the displacement current $-v\epsilon_0\left.\p{z}{\bf E}\right|_{z_{\rm tip}}$. However, the numerical simulations in figure \ref{fig:on_axis_currents} contradict this. In fact, we observe that at $z_{\rm tip}$ the displacement current vanishes since the electric field is maximal and conversely that $j_{e, \rm tip}$ does not vanish, which is in line with our reasoning. 

\subsection{Ionization and electric field in the charge layer}\label{sec:ionization_front}

As ions are essentially immobile within the propagating streamer head, the degree of ionization is best determined by the ion density $n_{i,\rm ch}$ behind the charge layer. An old approximation dating back to \cite{Dyakonov1989, Naidis1997} is 
\begin{equation}\label{eq:frontionization_planar}
    n_{i,\rm ch}^{\rm old ~approx} \approx \frac{\epsilon_0}{e}\int_0^{E_{\rm max}}\alpha_{\rm eff}(E)\,dE,
\end{equation}
where we use $E=|{\bf E}|$. In the appendix of \cite{Li2007DeviationsHeads} this equation is derived for planar negative streamer ionization fronts without electron diffusion or photo-ionization. The approximation is easily derived from the two following equations: equations (\ref{eq:pos_ion_transport}) and (\ref{eq:Si}) together yield 
\begin{equation} \label{eq:dtni}
    \partial_t n_i = |{\bf j}_e|\alpha_{\rm eff}(E),
\end{equation}
and equation (\ref{eq:conservationoftotalcurrent}) reads $\epsilon_0\partial_t E = -e j_e$, if the total current ahead of the charge layer vanishes. This is the case, if the electron density ahead of the planar front vanishes, and if the electric field ahead of the front does not change in time.

According to \cite{Li2007DeviationsHeads}, equation (\ref{eq:frontionization_planar}) is a good approximation of the numerical solutions of planar negative ionization fronts without photo-ionization in a time independent electric field; the error is only 5 to 10\%.
However, in simulations of positive curved streamer fronts with photo-ionization as shown in \cite{Luque2010SpritesVelocity, Li2022AMixtures}, the ionization density is about twice as high as given by the classical approximation \eqref{eq:frontionization_planar} (in particular, see table B1 of \cite{Li2022AMixtures}). In table \ref{tab:ionization_degree} we make a similar comparison and confirm the discrepancy of equation \eqref{eq:frontionization_planar} as an approximation of the ionization density of positive streamers.

\begin{table}[]
    \centering
    \caption{The ionization density $n_{i,\rm ch}$ \mbox{($\times 10^{19}$~m$^{-3}$)} for streamers in different background fields. All streamers are taken at \mbox{$\zeta_{\rm tip}=15$~mm}. We compare the old approximation (equation \eqref{eq:frontionization_planar})  and our approximation (equation \eqref{eq:frontionization_general}) with our simulated results.}
    \begin{tabular}{c|ccc}
         & classical & new & \\
         & eq.~\eqref{eq:frontionization_planar} & eq.~\eqref{eq:frontionization_general} & simulation \\
         \hline
         4.5 kV/cm & 11.9 & 21.9 & 25.6 \\
         10 kV/cm & 3.3 & 5.4 & 6.3 \\
         14 kV/cm & 3.4 & 5.6 & 6.3 \\
         24 kV/cm & 4.5 & 7.6 & 8.4 
    \end{tabular}
    \label{tab:ionization_degree}
\end{table}

A first hypothesis was that the approximation (\ref{eq:frontionization_planar}) only covers the part of the front where the electric field decays from its maximal value $E_{\rm max}$ to a low value inside the channel, and that it misses the avalanche zone ahead of the charge layer where the electric field increases to its maximum \cite{Luque2010SpritesVelocity}. This avalanche zone is essentially absent without background ionization and photo-ionization, but very present in air. However, the ionization created in the avalanche zone contributes relatively little ionization. We discuss this later in more detail in section \ref{sec:results} and figure \ref{fig:LE_log}.

We will now show that the total current density ${\bf j}_{\rm tot}$ from the avalanche zone into the curved charge layer contributes significantly to the ionization behind the front for positive streamers in air.  The derivation of the new approximation is analogous to the earlier one in \cite{Li2007DeviationsHeads}.
We start from (\ref{eq:dtni}) and express $j_e$ in terms of $j_{\rm tot}$
\begin{equation}
    \p{t} n_i =\frac{1}{e}\Big|{\bf j}_{\rm tot}-\epsilon_0\p{t}{\bf E}\Big|\alpha_{\rm eff}. \label{eq:intermediate_frontionization}
\end{equation}
This can be further simplified due for steady motion and because the  vectors ${\bf j}_{\rm tot}$ and ${\bf E}$ are parallel on the axis
\begin{equation}\label{eq:ni_equation_in_chargelayer}
    \p{z}n_i=\left(\frac{\epsilon_0}{e}\p{z}E+\frac{j_{\rm tot}}{ev}\right)\alpha_{\rm eff}.
\end{equation}
Integration through the ionization front gives
\begin{align}\label{eq:frontionization_general}
    n_i(z) &= n_{i,\rm tip} + \frac{\epsilon_0}{e}\int_{E(z)}^{E_{\rm max}}\alpha_{\rm eff}(E)\,dE \\
    & + \frac{1}{ev}\int_z^{z_{\rm tip}}\alpha_{\rm eff}(E(z))j_{\rm tot}\,dz. \nonumber
\end{align}
The first term is obtained after integration by substitution $(\partial_z E \:dz = dE)$. It reproduces the old approximation \eqref{eq:frontionization_planar}
when it is evaluated at $z=z_{\rm ch}$ and when $E(z_{\rm ch})$ is approximated as vanishing. 
The second term requires further analysis. We approximate $j_{\rm tot}$ by the constant $-ej_{e, \rm tip}$ according to equation \eqref{eq:J_dependence_on_r}. Furthermore, we need the spatial profile of $E(z)$ to evaluate $\alpha_{\rm eff}(E(z))$ under the integral.
Here we adopt a heuristic parametrization of $E$ and leave further analysis to future work. In figure \ref{fig:on_axis_currents} we see that within the layer the charge number densities $n_q$ have an approximately Gaussian profile that can be parameterized as
\begin{align}
    & n_q(z) = \frac{N_q}{\sigma\sqrt{2\pi}}\exp\left(-\frac{1}{2}\left(\frac{z-R}{\sigma}\right)^2\right), \label{eq:approx-nq} 
\end{align}
with
\begin{equation}
         N_q = \int_{z_{\rm ch}}^{z_{\rm tip}}n_q\;dz =\frac{\epsilon_0}e\;(E_{\rm max}-E_{\rm ch}), %\quad \sigma = \frac{\tilde\ell}2, 
     \label{eq:approx-Nq}
\end{equation}
an approximate normalization constant provided that $\sigma \ll \ell$.
    
Next, we use that over its small width the layer is only weakly curved, and we use a planar approximation \mbox{$\partial_z E = e n_q/\epsilon_0$} to calculate the electric field as $E(z)$ by integrating over $n_q$
\begin{equation}
     E(z) = E_{\rm max} - \frac e{\epsilon_0}\int_z^{z_{\rm tip}} n_q(z) dz. \label{eq:approx-E}
\end{equation}
This heuristic parametrization of the electric field is shown in the middle panels of figure \ref{fig:on_axis_currents} together with the results of the axisymmetric simulations. The parametrizations of $E$ are in agreement with the simulated results when we choose \mbox{$\sigma = \tilde\ell/3$} for the steady streamer and \mbox{$\sigma = \tilde\ell/2$} for the accelerating ones. Furthermore, we remark explicitly that equation \eqref{eq:approx-nq} is only used to motivate and evaluate the parameterization for $E$ in equation \eqref{eq:approx-E}. \\

Using equation \eqref{eq:approx-E} as an approximation for the electric field within the charge layer, we can calculate the ionization density by evaluating equation \eqref{eq:frontionization_general} at $z_{\rm ch}$. In table \ref{tab:ionization_degree} we compare this approximation, when all macroscopic parameters are extracted from simulations. We observe good agreement, with relative errors between 10-15\%.

 \subsection{Electron density in the charge layer}\label{sec:conservation_of_charge}

 Our derivation of the electron density within the charge layer starts from the fundamental equation of charge conservation
\begin{equation}
    e\p{t}n_q=-\nabla \cdot {\bf j}.
\end{equation}
Since we have uniform translation and a planar front we can write
\begin{equation}\label{eq:flat}
   v n_q=j_{e, \rm ch}-j_e,
\end{equation}
where $j_{e, \rm ch}$ has been introduced as an integration constant. As a side note, a similar relation has also been proposed in \cite{Naidis1997, Babaeva2021UniversalMedia}, but there the integration constant has been explicitly neglected. However, in figure \ref{fig:on_axis_currents} we see that $j_{e, \rm ch}$ and $j_{e, \rm tip}$ are significant. Continuing our derivation, we use \mbox{$n_q=n_i-n_e$} and
rearrange the terms in this equation such that we find an expression for the electron density profile in the charge layer
\begin{equation}\label{eq:ne_parameterization}
    n_e(z) = \frac{vn_i(z) - j_{e, \rm ch}}{v+v_{\rm dr}},
\end{equation}
with the charge drift velocity $v_{\rm dr}=\mu E$. (Note that electrons drift with $-v_{\rm dr}$). This determines $n_e(z)$ since $n_i(z)$ is given by equation \eqref{eq:frontionization_general}. By evaluating this expression at $z_{\rm ch}$ or $z_{\rm tip}$ and using equation \eqref{eq:eqv} we find quasi-neutrality: \mbox{$n_{e, \rm ch} = n_{i, \rm ch}$} and \mbox{$n_{e, \rm tip} \approx n_{i, \rm tip}$.} Note that the implied quasi-neutrality at $z_{\rm tip}$ only holds as an approximation, see figure \ref{fig:on_axis_currents}. \\

Moreover, integration of equation \eqref{eq:flat} through the charge layer and using \eqref{eq:approx-Nq} results in
\begin{equation}\label{eq:eql}
    \int_{z_{\rm ch}}^{z_{\rm tip}} e(j_{e, \rm tip}-j_e)\, dz = v\epsilon_0(E_{\rm max} - E_{\rm ch}).
\end{equation}
This can be interpreted as a physical connection between the movement of a positive charge layer (represented by a discontinuity in the electric field) and the separation of charge. The latter can be directly expressed by the electric current integrated through the charge layer. 

\section{The avalanche zone (\mbox{$z\geq z_{\rm tip}$)}\label{sec:avalanchezone}}

The avalanche zone is defined as the region ahead of the space charge layer where space charges can be neglected, and where the electric field is above the breakdown value.
This means that the electric field near this layer is dominated by the electric charges in the layer, and that charges in the avalanche zone move in this externally determined field, but do not contribute to it.

In the avalanche zone, different approximations have to be made than in the charge layer:
\begin{enumerate}[label=(\roman*)]
    \item As said above, the influence of the local charges on the electric field is negligible, $\nabla\cdot {\bf E}=0$, so the avalanche develops in an externally determined electric field.
    \item The dynamics inside the charge layer were described using the planar front approximation because \mbox{$\ell\ll R$}, but the planar front approximation is not valid in the avalanche zone. We therefore do account for the curvature of the charge layer in the avalanche zone. We do so by approximating the charge layer at the streamer tip as a hemisphere with a radius $R$, see figures \ref{fig:shape_and_grid} and \ref{fig:photoi_configuration}.
    \item Electron diffusion is still neglected but photoionization now needs to be included. Although the impact ionization is much stronger than the photoionization, the non-locality of the photoionization is essential to create seed electrons in the avalanche zone. 
\end{enumerate}

\subsection{Equation for electron density in the avalanche zone}

The drift-diffusion-reaction equation \eqref{eq:electron_transport} for the electron dynamics on the axis of the avalanche zone can be simplified as follows.
First we remark that with the approximations above and with the chain-rule we can write on the axis
\begin{align}
    \nabla\cdot (\mu n_e {\bf E}) & = {\bf E}\cdot\nabla(\mu n_e) + \mu n_e \nabla\cdot {\bf E},\nonumber\\
    &=E\p{z}{(\mu n_e)},\\
    &=v_{\rm dr}\p{z}{n_e} + \frac{\p{z}{\mu}}{\mu} v_{\rm dr} n_e,\nonumber
\end{align}
The electron dynamics of equation \eqref{eq:electron_transport} then becomes, in the comoving frame on the axis,
\begin{equation}\label{eq:electrondynamics_simplifiedequation}
    (v+v_{\rm dr})\p{z}n_e+\frac{\p{z}\mu}{\mu} v_{\rm dr}n_e + S_i+S_{ph}=0.
\end{equation}
In the next section we derive an expression for $S_{ph}$.

\subsection{Coupling between avalanche zone and charge layer}

The dynamics in the avalanche zone are coupled to the other discharge regions. More precisely, the charge layer together with the channel generate the enhanced electric field in the avalanche zone, and the charge layer also emits the large majority of photons that generate photoionization and initiate the ionization avalanches in the avalanche zone.

The electric field near the charge layer and near the streamer axis are approximated by a uniformly charged sphere 
\begin{equation}\label{eq:E_approx}
    E(z) = \frac{z_{\rm tip}^2 (E_{\rm max}-E_{\rm bg})}{z^2} + E_{\rm bg},
\end{equation}
as argued above. 

\begin{figure}
    \centering
    \includegraphics[width=0.45\textwidth]{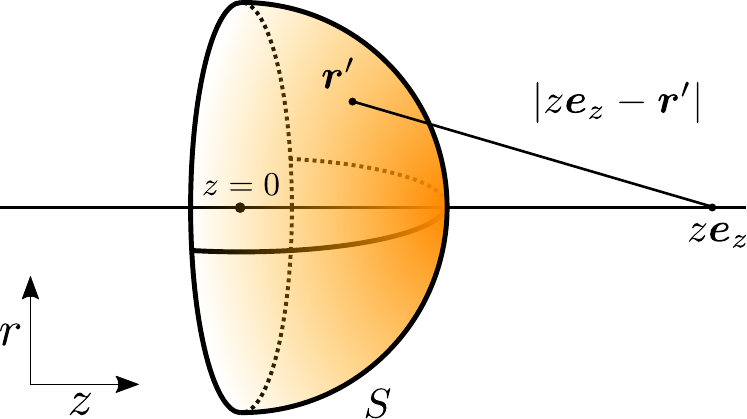}
    \caption{The configuration used for computing the photo-ionization source term. The charge layer is approximated by a hemisphere $S$ with radius $R$ centered around \mbox{$z=0$}. The color indicates that in reality the front is not radiating with uniform intensity but fades at the edges (even though we do not account for this here). Also shown is the path of a photon produced at $\bf r'$ and absorbed at $z{\bf e}_z$. Photoionization then creates electron avalanches that develop the local electric field. We use the avalanches on the $z$-axis for our approximations.}
    \label{fig:photoi_configuration}
\end{figure}

For photoionization in air, the photons are mainly produced in the charge layer, because the majority of high-energy collisions occurs here, as will be shown in figure \ref{fig:LE_log}. Photons originating from the avalanche zone are therefore neglected. Moreover, since typical absorption lengths (33$-$1900~$\mu$m for dry air at 1~bar and 300~K) are large compared to $\ell$, cf figure \ref{fig:vRln_comparison}, we can essentially treat the charge layer as a surface. Accordingly, we approximate equation \eqref{eq:sim_photoionizationsourceterm} by a surface integral
\begin{equation} \label{eq:Sph1}
    S_{ph}(z) = \iint_S \frac{I({{\bf r}'})f(|z{\bf e}_z-{\bf r'}|)}{4\pi |z{\bf e}_z-{\bf r'}|^2}d^2r',
\end{equation}
with ${\bf e}_z$ the unit vector in the $z$-direction, and the coordinates ${\bf r'}$ now lie on the surface $S$. For simplicity, we take $S$ to be the surface of a hemisphere with radius $R$ centered at \mbox{$z=0$}. This is illustrated in figure \ref{fig:photoi_configuration}.

The general photon source term from equation \eqref{eq:sim_photonsource} is now approximated as
\begin{equation}  \label{eq:Sph2}
    I({\bf r'}) = \mathcal{A}({\bf r'})I^*,
\end{equation}
with $I^*$ the surface density of photon production
\begin{equation}  \label{eq:Sph3}
    I^* = \frac{p_q}{p+p_q}\xi v n_{i,\rm ch}
\end{equation}
on the streamer axis. Here ${p_q}/(p+p_q)$ is the quenching factor of the photon emitting state. The excitation of the photon emitting state is approximated as impact ionization $S_i$ times a proportionality factor $\xi$. Note that the impact ionization has to be integrated over the width of the charge layer \mbox{$\int S_i\,dz=v(n_{i,\rm ch} - n_{i,\rm tip})$} which is obtained after integrating \mbox{$-v\partial_z n_i = S_i$} (from equation \eqref{eq:pos_ion_transport}) across the charge layer. Finally, since \mbox{$n_{i,\rm tip}\ll n_{i,\rm ch}$} we have omitted the dependency on $n_{i,\rm tip}$.

$\mathcal{A}({\bf r'})$ is a function that can account for the fact that the impact ionization and thus the photon radiation in the charge layer diminishes in the off-axis direction. However, for simplicity we take \mbox{$\mathcal{A}({\bf r'}) = 1$}. Naturally this will slightly overestimate photon radiation. 

\subsection{Solving the electron density in the avalanche zone} \label{sec:solving_electron_density}

We will now solve equation \eqref{eq:electrondynamics_simplifiedequation}. To do so we first introduce the short hand notation
\begin{equation}\label{eq:electrondynamics_shorthand}
    \p{z}n_e + \lambda(z)n_e = -K(z),
\end{equation}
with $\lambda(z)$ the electron avalanche growth function
\begin{equation}
    \lambda(z) = \frac{v_{\rm dr}(E(z))}{v+v_{\rm dr}(E(z))}\left(\alpha_{\rm eff}(E(z))+\frac{\p{z}{\mu}}{\mu}\right),
\end{equation}
and $K(z)$ the photoelectron source term
\begin{equation}\label{eq:K}
   K(z) = \frac{S_{ph}(z)}{v+v_{\rm dr}(E(z))}, 
\end{equation}
in the external electric field $E(z)$ from equation \eref{eq:E_approx}. $S_{ph}(z)$ is determined by equations \eref{eq:Sph1} -- \eref{eq:Sph3} as 
a surface-integral corresponding to the parametrized charge layer. For given $I^*$, equation \eqref{eq:electrondynamics_shorthand} is an ordinary differential equation for $n_e$ that is solved as
\begin{equation}\label{eq:electrondensity_generalsolution}
    n_e(z) = \int_{z}^\infty  K(y)\;e^{\int_{z}^y \lambda(x)\;dx}\;dy.
\end{equation}
This solution can be interpreted as a superposition of electron avalanches. The electron avalanches are continuously created by a photoelectron density $K$. The avalanches grow in the electric field as described by $\lambda$ which contains the effects of impact ionization $\alpha_{\rm eff}$ and of electron mobility $\mu(E)$.

For further evaluation, it is interesting to discuss the structure of this solution and the implications for the electron and ion densities at the front and back end of the charge layer, $z_{\rm tip}$ and $z_{\rm ch}$. We find that equation \eqref{eq:electrondensity_generalsolution} can be rewritten as
\begin{equation}\label{eq:newphotobalance}
    \frac{n_{e,\rm tip}}{n_{i,\rm ch}} = F(v,R,E_{\rm max},E_{\rm bg}),
\end{equation}
with an explicit equation for the function $F$ that does not depend on any electron or ion densities. Here $R$, $E_{\rm max}$ and $E_{\rm bg}$ determine the electric field $E(z)$ in the avalanche zone according to \eref{eq:E_approx}. That $F$ does not depend on the particle densities, is due to the linear nature of the avalanche zone without local space charge effects: twice as many photons emitted from the charge layer will create twice as many avalanches and twice as many electrons arriving at $z_{\rm tip}$ which in turn emit twice as many photons from the charge layer. 

The explicit equation for the function $F$ is
\begin{align}\label{eq:F}
    F & = \frac{p_q}{p+p_q}\;\xi \; \int_{z_{\rm tip}}^\infty dy\; 
    \frac v{v+v_{\rm dr}(y)} \;e^{\int_{z_{\rm tip}}^y \lambda(x)\;dx}
    \nonumber\\
    & \qquad \cdot \iint_S d^2r'\;\mathcal{A}({\bf r'})
    \;\frac{f(|y{\bf e}_z-{\bf r'}|)}{4\pi |y{\bf e}_z-{\bf r'}|^2} ,
\end{align}
where the first line contains the field dependent electron dynamics on the streamer axis, and the second line the field independent photon dynamics between the charge layer and the axis. 

An analysis of the avalanche zone along similar lines was proposed in \cite{Pancheshnyi2001}, but they only account for photons produced in the avalanche zone and neglect the contribution from the charge layer. However figure \ref{fig:on_axis_currents} shows that ionization in the charge layer, and therefore the associated photon production, is far more important. In our approach we do take the charge layer as the dominant photon source. The same reasoning was also given in \cite{Lehtinen2021, Pavan2020}. In addition to this we have derived an improved photoionization balance on the basis of consistent electrodynamics in the charge layer and avalanche zone, equation \eqref{eq:newphotobalance}. This formula replaces the photoionization balance proposed in \cite{Pancheshnyi2001}. We finally remark that the balance between the dynamics of photons and of electron avalanches resembles a self-sustained DC discharge, with the difference that the anode is replaced by a propagating streamer head with self-consistent shape. 

\section{The electrostatic field and the head potential}\label{sec:head_potential}

\subsection{Streamer head potential}
As recalled in~\cite{Nijdam2020ThePhenomena}, the electrostatic approximation for the electric field ${\bf E}=-\nabla\varphi$ is sufficient for streamer physics. Therefore the line integral between any two points is independent of the path taken between them
\begin{equation}\label{eq:fundamentalgradient}
    \int_{\mathcal{C}} {\bf E}\cdot dl = \phi({\bf r}) - \phi({\bf r'}),
\end{equation}
with $\mathcal{C}$ any continuous curve which starts at ${\bf r}$ and ends at ${\bf r'}$. This concept will be applied to derive a relation between the electrostatic properties of the channel and the head. \\

We shall use equation \eqref{eq:fundamentalgradient} to solve two path-integrals, the first corresponding only to the background field and the second to the field with a streamer present. In both cases $\mathcal{C}$ equals the axis of propagation, i.e.\ $\zeta$-axis, which gives \mbox{${\bf r}=0$} and ${\bf r'}$ on the opposing electrode. For the streamers in this work $\zeta_{\rm tip}$ is far away from the opposing electrode, which means that boundary effects are negligible and we can take ${\bf r'}$ at infinity. When we subtract the two integrals we find
\begin{equation}\label{eq:conservationofpotential}
    \int_{0}^\infty \big(E(\zeta)-E_{\rm bg}\big)\,d\zeta = 0,
\end{equation}
since the potential at ${\bf r}$ and ${\bf r'}$ is the same and therefore the right-hand side vanishes. This fundamental property has been considered by previous authors \cite{Kulikovsky1998AnalyticalCalculations, Pancheshnyi2001, Guo2022AAir, Li2022AMixtures, Starikovskiy2022TheGap}. Equation \eqref{eq:conservationofpotential} will be split in two intervals with different dynamics, namely: the streamer channel $[0,\zeta_{\rm tip}]$ and the avalanche zone $[\zeta_{\rm tip}, \infty)$. We shall treat each of these intervals separately.

\subsubsection{Potential across the channel. $\quad$ } The potential across the channel requires different treatment for steady and accelerating streamers.\\

For a steady streamer the channel electric field decays back to the background field. In general the profile of the channel electric field is determined by currents in the streamer channel \cite{Luque2014GrowingStreamers, Luque2017StreamerChannels}. For now, modelling the charge distribution within the channel is not considered. Instead, we suggest a plausible channel electric field profile for steady streamers. In section \ref{sec:definitions_and_conventions} we have discussed how dynamics in the channel are related to an electron loss time scale $\tau$, which in turn defines an electron loss length $L_{\rm loss}$.  We use these concepts to impose
\begin{align}
    E(\zeta) &= E_{\rm bg} + (E_{\rm ch} - E_{\rm bg})\exp\left(\frac{\zeta - \zeta_{\rm ch}}{L_{\rm loss}}\right),\nonumber\\
    &\text{for } \zeta<\zeta_{\rm ch}.
\end{align}
Substituting this into equation \eqref{eq:conservationofpotential} results in
\begin{equation}
    \int_{0}^{\zeta_{\rm ch}} \big(E(\zeta)-E_{\rm bg}\big)\,d\zeta = L_{\rm loss}\left(E_{\rm bg} - E_{\rm ch}\right).
\end{equation}

For the accelerating streamers considered in this work we have $L\ll L_{\rm loss}$, which means it is more reasonable to work with an averaged channel electric field $\bar E_{\rm ch}$. By holding that \mbox{$E_{\rm ch} = \bar E_{\rm ch}$} over the length of the channel we can obtain a similar result
\begin{equation}
    \int_{0}^{\zeta_{\rm ch}} \big(E(\zeta)-E_{\rm bg}\big)\,d\zeta = L\left(E_{\rm bg} - \bar E_{\rm ch}\right).
\end{equation}

\subsubsection{Potential across the avalanche zone. $\quad$ } 
In the avalanche zone the electric field was approximated by that of a uniformly charged sphere, equation \eqref{eq:E_approx}. Using this the potential across the avalanche zone simplifies approximately to
\begin{equation}\label{eq:headpotential}
    \int_{\zeta_{\rm tip}}^{\infty} \big(E(\zeta)-E_{\rm bg}\big)\,d\zeta = R(E_{\rm max}-E_{\rm bg}).
\end{equation}
This gives the final result
\begin{equation}\label{eq:eqR}
    R(E_{\rm max}-E_{\rm bg}) = L\left(E_{\rm bg} - E_{\rm ch}\right).
\end{equation}
To keep notation simple we have no longer discerned between $L_{\rm loss}$ or $\bar E_{\rm ch}$ for the separate cases of steady and accelerating streamers.

\section{Solving the approximations}\label{sec:solutionandresults}

\subsection{Solution method}\label{sec:implementation}
We now assume that velocity $v$, radius of curvature $R$, length $L$ and background electric field $E_{\rm bg}$ are given, for example by experimental measurements, and we estimate four unknowns that are much more difficult to measure: ionization density $n_{i,\rm ch}$, maximal electric field $E_{\rm max}$, channel field $E_{\rm ch}$, and charge layer width $\ell$. To that end we shall formulate a system of four relations from which these unknowns will be determined. \\

In the previous sections we have derived equations \eqref{eq:eqv} and \eqref{eq:eql} by analyzing the dynamics of the charge layer \mbox{$z_{\rm ch}\leq z<z_{\rm tip}$}, where \mbox{$z_{\rm tip, ch}=R\pm\ell/2$}. These are the first and second relations. On the basis of electrostatics we have related the head potential to the streamer length in equation \eqref{eq:eqR}, which is the third relation. Finally, we require that charge layer and avalanche zone electron dynamics are consistent (cf.\ section \ref{sec:solving_electron_density}). This introduces the last relation, namely equation \eqref{eq:newphotobalance}. For convenience, we repeat our relations here
\begin{align}
     j_{e, \rm ch} &= j_{e, \rm tip}, \label{eq:recap_start}\\
     v\epsilon_0(E_{\rm max}-E_{\rm ch}) &= \int_{z_{\rm ch}}^{z_{\rm tip}} e(j_{e, \rm tip}-j_e)\,dz, \\
     R(E_{\rm max}-E_{\rm bg}) &= L\left(E_{\rm bg} - E_{\rm ch}\right),\\
     \frac{n_{e,\rm tip}}{n_{i,\rm ch}} &= F(v,R,E_{\rm max},E_{\rm bg}). \label{eq:recap_stop}
\end{align}
The function $F$ is defined in equation \eqref{eq:F} and the electron current density is defined as \mbox{$j_{e}=-\mu n_e E$}. The above system of equations has $8$ independent parameters:
\begin{equation}\label{eq:independent_parameters}
v,\ R,\ L,\ E_{\rm bg},\ n_{i,ch},\ E_{\rm max},\ E_{\rm ch} \text{ and } \ell.    
\end{equation}
All other quantities are determined by these $8$ parameters. To see this, we summarize our approximations in the two regions:
\begin{itemize}
\item  In the avalanche zone \mbox{$(z\geq z_{\rm tip})$} the electric field and the electron density are approximated by (equations \eqref{eq:E_approx} and \eqref{eq:electrondensity_generalsolution})
\begin{align}
    E(z) &= \frac{z_{\rm tip}^2 (E_{\rm max}-E_{\rm bg})}{z^2} + E_{\rm bg}, \\
    n_e(z) &= \int_{z}^\infty  K(y)\;e^{\int_{z}^y \lambda(x)\;dx}\;dy,\\
    & \text{for } z\geq z_{\rm tip}.\nonumber   
\end{align}
Notably, the function $K(y)$ (equation \eqref{eq:K}) accounts for the production of photoelectrons and is proportional to $n_{i, ch}$. The electron density and the electric field by definition give $j_{e}$ and therefore $j_{e, \rm tip}$. Finally, we have assumed that space charge effects are negligible in the entire avalanche zone. We therefore also assume quasi-neutrality at the tip \mbox{$n_{i, \rm tip}\approx n_{e, \rm tip}$}.  

\item In the charge layer \mbox{$(z_{\rm ch}\leq z<z_{\rm tip})$} the electric field and densities are approximated by (equations \eqref{eq:approx-E}, \eqref{eq:frontionization_general}, \eqref{eq:ne_parameterization})
\begin{align}
    E(z) &= E_{\rm max} - \frac e{\epsilon_0}\int_z^{z_{\rm tip}} n_q(z) dz,\label{eq:E_recap}\\
    n_i(z) &= n_{i,\rm tip} + \frac{\epsilon_0}{e}\int_{E(z)}^{E_{\rm max}}\alpha_{\rm eff}(E)\,dE, \nonumber\\
    & + \frac{1}{ev}\int_z^{z_{\rm tip}}\alpha_{\rm eff}(E(z))j_{\rm tot}\,dz,\\
    n_e(z) &= \frac{vn_i(z) - j_{e, \rm ch}}{v+v_{\rm dr}},\\
    & \text{for } z_{\rm ch}\leq z< z_{\rm tip},\nonumber
\end{align}
where $n_q$ in equation \eqref{eq:E_recap} is a parametrization given in equation \eqref{eq:approx-nq}. The quantities $n_{i, \rm tip}$ and $j_{e, \rm tip}$ are determined by the avalanche zone. Quasi-neutrality in the channel gives \mbox{$n_{i, \rm ch}= n_{e, \rm ch}$}. Thus we can evaluate $j_{\rm tot}$ and $j_e$ within the charge layer.   
\end{itemize}

The objective is then to determine $4$ parameters in \eqref{eq:independent_parameters}, since we consider that \mbox{$(v,\ R,\ L, E_{\rm bg})$} are fixed by observations. The remaining four, which we call \mbox{$m=(n_{i,\rm ch},\, E_{\rm max},\, E_{\rm ch},\, \ell)$}, have to satisfy our relations \eqref{eq:recap_start}-\eqref{eq:recap_stop}. Solving this system of equations is equivalent to finding the roots of the four-dimensional vector-function $\mathcal{S}$, which is defined as the difference between the left- and right-hand sides of equations \eqref{eq:recap_start}-\eqref{eq:recap_stop}. Thus $m$ is a consistent solution if it satisfies
\begin{equation}\label{eq:streamer_root_equation}
    \mathcal{S}(m) = 0.
\end{equation}
Due to the complexity of $\mathcal{S}$ we employ an iterative root-finding algorithm that solves equation \eqref{eq:streamer_root_equation} using a modification of the Levenberg-Marquardt algorithm \cite{more1980user}. Such an algorithm starts from an initial guess $m^0$ and produces a sequence of values $m^k$ that converges to the root. We emphasize again that the input parameters $v$, $R$, $L$ and $E_{\rm bg}$ in addition to $m^k$ are sufficient to evaluate $\mathcal{S}(m^k)$. Moreover, changing the initial guess seems to have no effect on the obtained solution $m$, suggesting that the solution $m$ is unique. We observe the same in numerical simulations \cite{Li2022AMixtures, Guo2022AAir}.

\subsection{Results}\label{sec:results}
\subsubsection{Steady streamer: }
\begin{figure}
    \centering
    \includegraphics[width=0.5\textwidth]{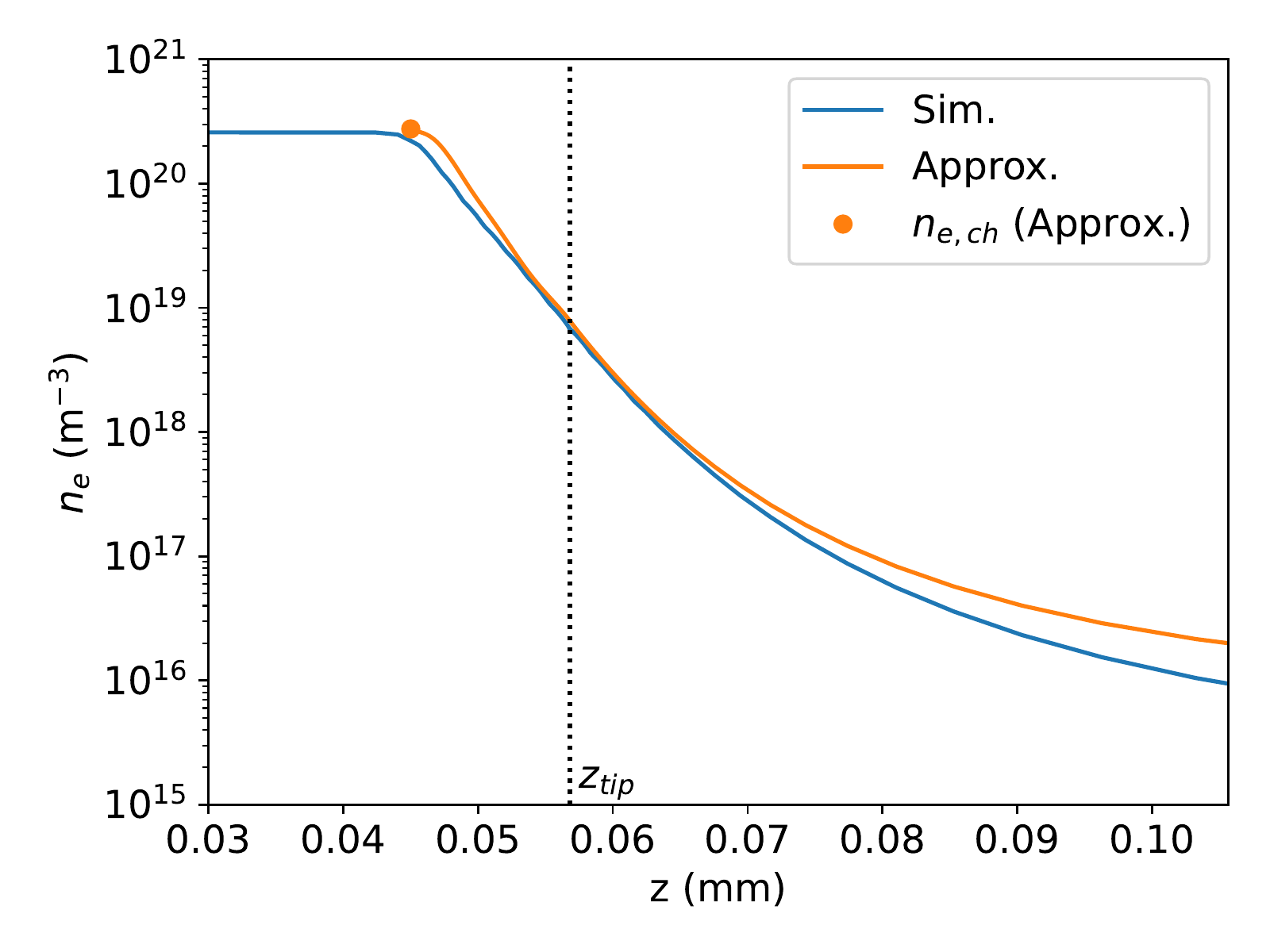}
    \caption{Our approximation (orange) for the electron density compared to numerical results (blue) of a steady streamer simulation. The applied background field is $4.5$~kV/cm. The approximated parameters used to make this comparison are evaluated in figure \eqref{fig:vRln_comparison}}
    \label{fig:LE_log}
\end{figure}

\begin{figure*}
    \centering
    \includegraphics[width=0.48\textwidth]{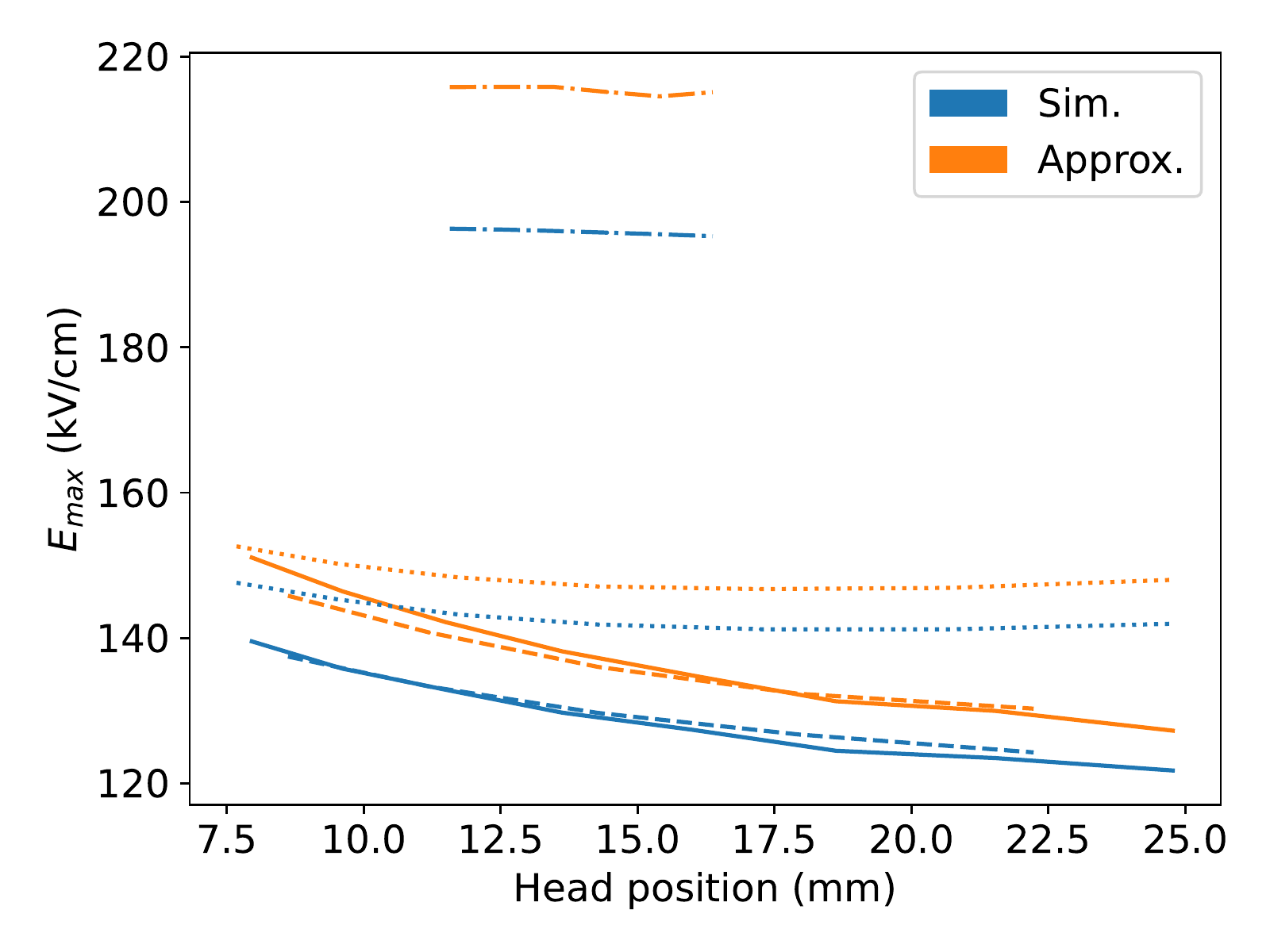}
    \includegraphics[width=0.48\textwidth]{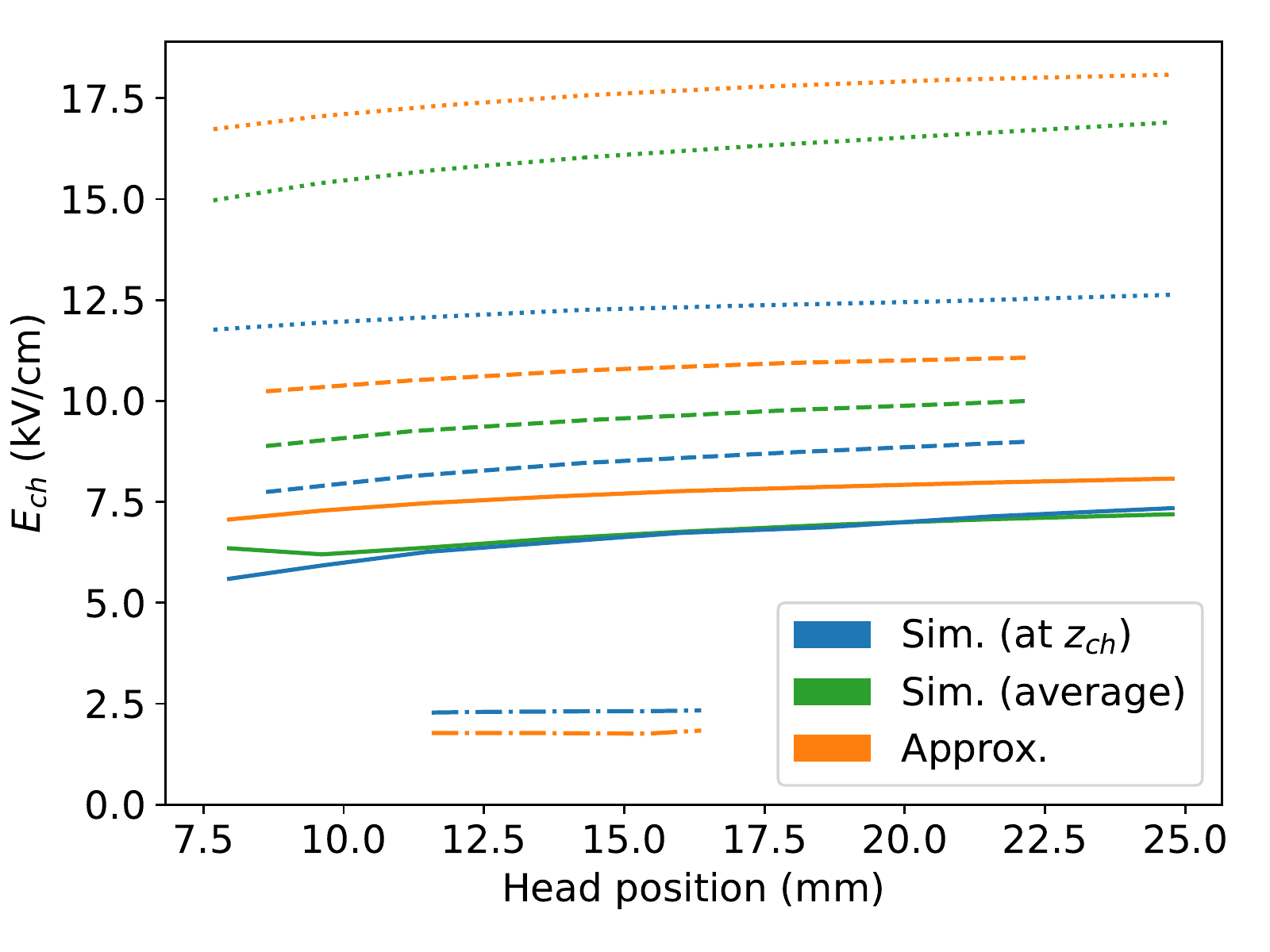}
    \includegraphics[width=0.48\textwidth]{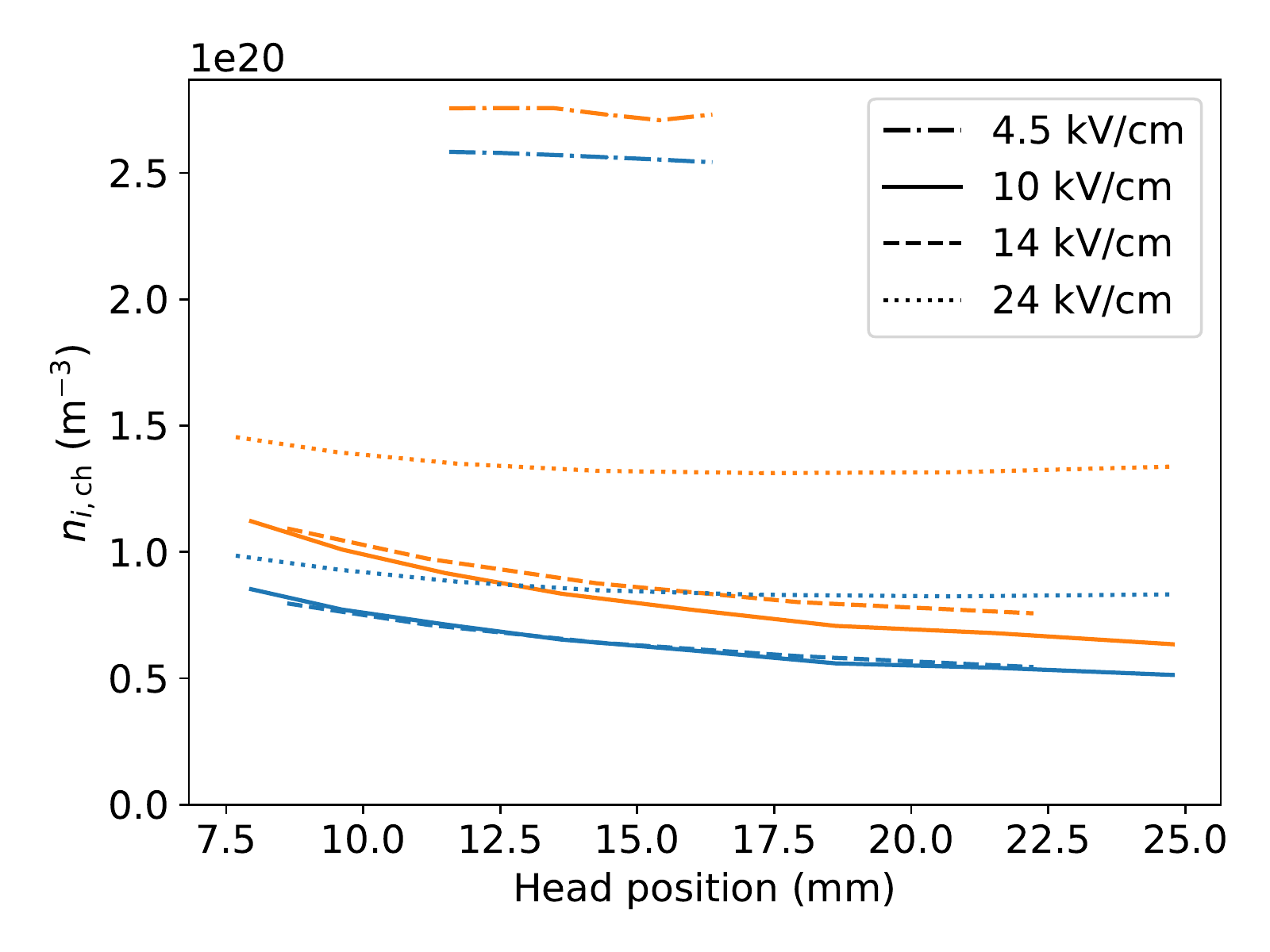}
    \includegraphics[width=0.48\textwidth]{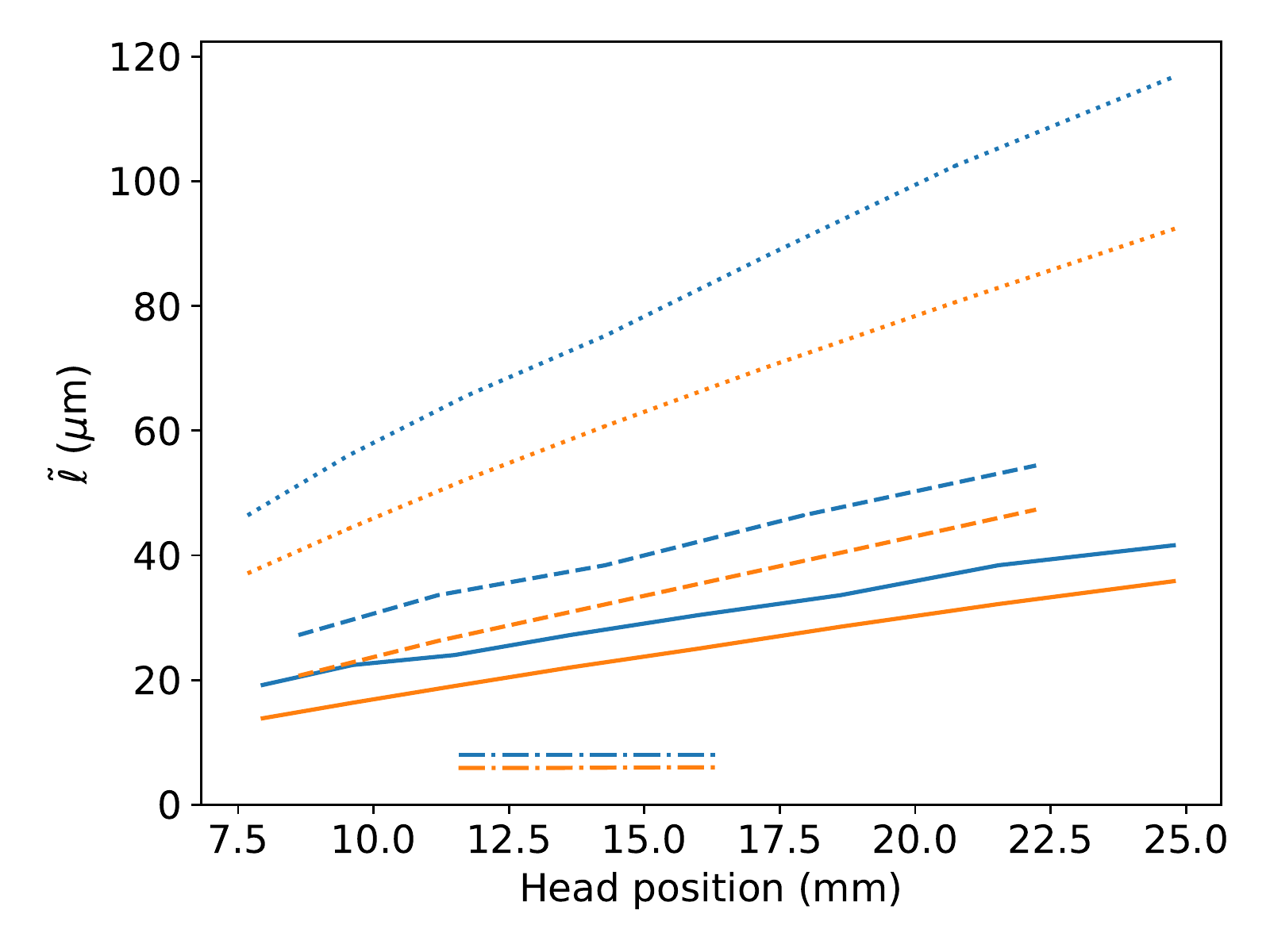}
    \caption{Comparison of simulations (blue, green) and our approximations (orange) for streamers with varying head positions in different background fields $E_{\rm bg}$. $E_{\rm bg}$, $L$, $R$ and $v$ were taken from the simulations and used to calculate the plotted approximations from (\ref{eq:streamer_root_equation}). The plotted quantities are the maximum electric field $E_{\rm max}$, the (average) channel electric fields $E_{\rm ch}$, the degree of ionization $n_{i,\rm ch}$ and the charge layer width $\tilde \ell$. The four background electric fields $E_{\rm bg}$ are plotted as \rule[.5ex]{0.5em}{.4pt}\,$\cdot$\,\rule[.5ex]{0.5em}{.4pt}\,$\cdot$ for $4.5$\,kV/cm (steady),
    \rule[.5ex]{2em}{.4pt} for $10$~kV/cm, \rule[.5ex]{0.5em}{.4pt}\,\rule[.5ex]{0.5em}{.4pt}\,\rule[.5ex]{0.5em}{.4pt} for $14$~kV/cm, and \mbox{$\cdot\,\cdot\,\cdot$} for $24$~kV/cm).}
    \label{fig:vRln_comparison}
\end{figure*}

In this section we will compare the approximated $n_e(z)$, $n_{i,\rm ch}$, $E_{\rm max}$, $E_{\rm ch}$ and $\tilde \ell$ with numerical simulations. We shall first do this comparison for the steady streamer.  To obtain these results we extracted \mbox{$E_{\rm bg}=4.5$~kV/cm}, \mbox{$v=0.076$~mm/ns}, \mbox{$R=49$~$\mu$m} and \mbox{$L_{\rm loss}=3.8$~mm} from simulation (see figure \eqref{fig:vandR}) and used these to solve equation \eqref{eq:streamer_root_equation}. \\

In figure \ref{fig:LE_log} we show our approximation for the axial electron density of the steady streamer (equations \eqref{eq:ne_parameterization} and \eqref{eq:electrondensity_generalsolution}). The approximated electron density was overlaid onto the results from the numerical simulation such that the respective $z_{\rm tip}$ overlap. We observe that our analytic formulae for the electron density profile in the avalanche zone reproduces the profile obtained from simulation well. In this figure we can also observe that more than $95\%$ of the ionization occurs in the charge layer. This underlines our earlier arguments that ionization predominantly occurs in the charge layer and that photons originating from the avalanche zone can be neglected. \\

The approximated parameters $n_{i,\rm ch}$, $E_{\rm max}$, $E_{\rm ch}$ and $\tilde \ell$ that were derived in this evaluation are shown in figure \ref{fig:vRln_comparison}. We observe good agreement with a maximum relative error of about $30\%$ for the prediction of $\tilde\ell$. The other parameters agree within $25\%$. 

\subsubsection{Accelerating streamers: }
As discussed in the introduction, we shall now apply our analysis developed for steady streamers to accelerating streamers. We include results, calculated in the same manner, for streamers at background electric fields of $10$, $14$ and $24$~kV/cm. The corresponding velocity and radius as a function of streamer length were already shown in figure \ref{fig:vandR}.\\

The approximated parameters are included in figure \ref{fig:vRln_comparison}. In this case we also observe good agreement with relative errors of at most several tens of percent. Only at $24$~kV/cm do we have relative errors of about \mbox{$50-60\%$} for the estimation of $n_{i,\rm ch}$. Furthermore, we also illustrate the error introduced by our simplified treatment of the channel electric fields. For accelerating streamers we have included both $E_{\rm ch}$ and the averaged $\bar E_{\rm ch}$ in figure \ref{fig:vRln_comparison}. In section \ref{sec:head_potential} we have used \mbox{$E_{\rm ch}=\bar E_{\rm ch}$} in order to obtain an equation for the channel electric fields without resolving the entire charge transport dynamics of the channel. However, this approximation is generally not true and the accuracy is worst for the $24$~kV/cm case. This has various causes, such as a persisting neutral seed (i.e.\ due to shorter propagation times the influence of initial conditions still persist), actual inhomogeneities in the channel or the influence of boundary conditions. \\

Overall, our model is also able to estimate the properties of streamers in higher background fields. Evidently, approximating the charge layers of accelerating streamer heads as planar fronts in a steady state gives reasonable results.

\section{Summary and outlook}
\subsection{Summary}
In this work we have proposed a model that characterizes a single positive air streamer on the basis of observable parameters. Overall, our approximations exhibit good agreement with numerical simulations of a steady streamer with typical relative errors below $30\%$. For accelerating streamers the errors are slightly higher, with a maximum deviation up to $60$\% in the highest considered background field.\\

\noindent Our most important theoretical contributions are:
\begin{itemize}
    \item We have constructed a self-contained axial model that can approximate macroscopic properties of steady streamer heads. This model also gives good results for accelerating streamer heads.
    \item We have shown how the quantities $n_{i,\rm ch}$, $E_{\rm max}$, $E_{\rm ch}$ and $\ell$ can be determined from the more easily observable parameters $R$, $v$, $L$ and $E_{\rm bg}$.
    \item We have provided a formula for the ionization density of a streamer. Notably this formula contains the contribution due to a non-zero total current density and is about twice as high as the classical formula.
    \item We have given a self-consistent description of electron dynamics which includes the implicit contribution due to photoelectrons produced in the avalanche zone. 
\end{itemize}

\subsection{Outlook}
For future work we recommend three possible improvements:
\begin{itemize}
    \item We have not considered explicitly solving the dynamics of the charge layer. Instead we have accounted for these dynamics by heuristic parameterizations. However, a numerical approach that resolves densities and the electric field inside the charge layer can be expected to improve the accuracy. Moreover such an approach could replace a number of parameterizations, which would lead to a more precise representation of streamer dynamics. 
    \item We have used two approaches for the channel electric fields. For accelerating streamers we have used an average value \mbox{$\bar E_{\rm ch} = E_{\rm ch}$}, and for steady streamers we have used an exponential decay with a prescribed length scale $L_{\rm loss}$. These clearly have their limitations. In future work we aim to combine the insights obtained in this research with models that explicitly evaluate the dynamics of the streamer channel, such as \cite{Luque2017StreamerChannels}. 
    \item All derivations in this work assume that the dynamics of the charge layer can be approximated in a planar front setting, since the dimensionless parameter $\ell/R$ is typically small. A systematic expansion in terms of $\ell/R$ will likely improve the accuracy of our model.
\end{itemize}

\noindent Finally we comment on the significance of our work regarding the development of accurate streamer tree models such as \cite{Luque2014GrowingStreamers, Luque2017StreamerChannels}. The current limitation of these models is that they lack a self-consistent description of velocity and radius of a streamer. These parameters are often imposed. However, our model can be combined with a tree model in order to overcome this critical limitation for positive streamers. 

\section*{Acknowledgements}
HF was funded by the European Union’s Horizon 2020 Research and Innovation Programme under the Marie Skłodowska-Curie Grant Agreement SAINT722337.

\section*{References}
\bibliography{References, references_manual}

\end{document}